\documentclass[journal]{IEEEtran}

\usepackage{amsmath,amssymb,amsfonts}
\usepackage[T1]{fontenc}
\usepackage[utf8]{inputenc}
\usepackage{graphicx}
\usepackage[export]{adjustbox}
\usepackage{booktabs}
\usepackage{multirow}
\usepackage{array}
\usepackage{float}
\usepackage{cite}
\usepackage{url}
\usepackage[unicode]{hyperref}
\graphicspath{{mathpix_assets/}}
\DeclareUnicodeCharacter{2713}{\checkmark}
\DeclareUnicodeCharacter{00D7}{\ensuremath{\times}}

\title{Multiple Vehicles and Traction Network Interaction System Stability Analysis and Oscillation Responsibility Identification}
\author{Xiangyu Meng, Qiao Zhang, Zhigang Liu, Guiyang Hu, Fang Liu, Guinan Zhang}

\begin{document}
\maketitle

\begin{abstract}
The electrical incompatibility between vehicles and traction network in railway system can result in system instability and oscillation overvoltage issues. To analyze the system stability, impedance-based frequency-domain methods are commonly used. However, the current impedance-based modeling methods face challenges in practical implementation due to the requirement of precise analytical models and detailed internal parameters for all vehicles. Moreover, multiple vehicles operate simultaneously in railway systems, each with different operating conditions and internal parameters, thereby influencing system stability to different extents. Therefore, it is crucial to accurately identify the critical vehicles to prevent resonance accidents. To address these challenges, a component connection-based modeling approach for the railway vehicle-grid system is proposed, which only requires the measured impedance results without the internal information of vehicles. In addition, a multilevel sensitivity analysis method is introduced to quantitatively identify the critical vehicles and internal parameters that influence system stability, which outperforms traditional sensitivity analysis methods in computational complexity. Furthermore, a system-level electrical compatibility test process for the railway vehicle-grid system is provided, incorporating the proposed stability and sensitivity analysis methods. Finally, case studies based on the real-world train schedule of a multivehicle-accessed railway vehicle-grid system are designed to verify the correctness of the proposed method.
\end{abstract}

Index Terms-Compatibility test, oscillation, railway, sensitivity analysis, stability, vehicle-grid system.

\section*{Nomenclature}
HSR High-speed railway.\\
LFO Low-frequency oscillation.

\begin{center}
\begin{tabular}{|l|l|}
\hline
HIS & Harmonic instability. \\
\hline
PoC & Point of connection. \\
\hline
PCC & Point of common coupling. \\
\hline
CCM & Component connection method. \\
\hline
EMUs & Electric multiple units. \\
\hline
VSC & Voltage source converters. \\
\hline
CW & Contact wire. \\
\hline
RW & Rail wire. \\
\hline
FW & Feeder wire. \\
\hline
AT & Auto transformer. \\
\hline
MFD & Mirror frequency decouple. \\
\hline
MIMO & Multiple-input multiple-output. \\
\hline
GNSC & Generalized Nyquist stability criterion. \\
\hline
TM & Traction mode. \\
\hline
POM & Preparation operation mode. \\
\hline
RBM & Regenerative braking mode. \\
\hline
OCLs & Overhead contact lines. \\
\hline
HIL & Hardware-in-the-loop. \\
\hline
\end{tabular}
\end{center}

\section*{I. Introduction}
BY THE end of 2021, China's HSR had reached an operating mileage of 40000 km , accounting for over two-thirds of the world's HSR mileage [1]. However, a series of HIS issues have occurred in the railway system in recent years. For instance, in 2015, LFO issues were discovered in the traction network voltage of a railway substation in Xuzhou, China, where multiple HXD2B locomotives were locked down due to the transient fluctuated line-side overvoltage and overcurrent at $0.6-2 \mathrm{~Hz}$ [2].

The railway HIS problems are caused by the interaction between vehicles and the traction power supply system, as defined in the EN 50388-2022 Standard [3]. To evaluate the harmonic risk associated with integrating new elements into railway vehicle-grid systems, compatibility tests are necessary. The railway infrastructure manager needs to perform systemlevel stability assessments, in collaboration with the rolling stock suppliers, to ensure the integration is free from hazards. To prevent HIS issues, the line-side controllers of the traction converter are required to tune passive properties within a given frequency range, where the real part of the vehicle's input admittance is positive. In recent years, numerous input-impedance or input-admittance models have been established for different\\[0pt]
types of vehicles to optimize the control parameters of traction converters based on the design requirement. Furthermore, the stability of the vehicle-grid system can be analyzed using frequency-domain stability analysis methods or electromagnetic transient simulation platforms. These methods unveil some of the HIS mechanisms in railway vehicle-grid systems. The negative impedance property of the traction converter can lead to HIS issues when connected to a weak traction power supply grid. However, many of the existing studies focus solely on the integration of vehicles and the traction grid at a single PoC, such as a grid with a single vehicle [4], [5] or multiple parallel vehicles [6]. These studies overlook the line impedance or AT impedance among different operating vehicles. To analyze the dynamic movement characteristic of vehicles and the inherent impedance characteristic of traction networks on system harmonic stability, some research has applied the eigenvalue-based criteria to modular state-space modeling at multiple PoCs [7], or utilized the zero-pole analysis to the impedance model of multiple vehicles accessed vehicle-grid system [8]. However, these methods require preknown control parameters or circuit structures of vehicles, which is typically unavailable due to intellectual property protection.

When multiple vehicles are connected to the traction power supply system, various factors can influence system stability. These factors include not only the electrical parameters of the vehicles (such as operating power and modes) but also the internal control parameters. However, the sensitivity of system stability to different factors varies. Identifying the most sensitive factor allows for significant improvements in system stability by adjusting or redesigning that particular factor, thereby facilitating system-level stability optimization. Classical time-domain sensitivity analysis has been widely employed to assess node voltage stability indices in synchronous generator-dominated power systems, where the correlation between state variables and oscillatory modes is described by the time-domain statespace model [9]. Nevertheless, this method has limitations when applied to system-level stability analysis of power electronicdominated power systems due to the complex computation processes involved in high-dimensional matrices. Moreover, it is not applicable to black-box systems, where the internal workings are unknown. Recently, sensitivity analysis methods have been developed for power systems with multiple converters based on frequency-domain impedance models. In [10], a frequencydomain sensitivity analysis method was proposed based on the residual polar relationship of the closed-loop transfer function. This method derives the sensitivity of system oscillation modes to characterize the effect of node voltage or branch current on system stability. However, this method requires knowledge of frequency-domain transfer functions to calculate poles and residuals, which are typically unavailable for black-box systems. In addition, sensitivity analysis based on minor loop gain of open-loop transfer function matrices was developed. This approach provides insights into the effect of different port voltage and current signals [11], or internal converter impedance on system stability [12]. However, further-level sensitivity analysis and implementation of this method to system-level design require further study. Moreover, unlike renewable energy\\
integrated power systems, the railway vehicle-grid system possesses challenges, such as vehicle movement and changes in operating points, making impossible direct application of these methods.

This study analyzes the railway vehicle-grid system under the all-parallel double-way feeding AT mode power supply topology, primarily due to the unique challenges caused by the railways in Western China's mountainous regions. These areas have continuous sections of long, steep slopes, which may result in the electrical phase separation being set on the ramp. Notably, "dead zones" within the electrical phase separation areas can lead to significant power and speed losses in trains, significantly impacting train performance. To deal with this problem, the double-way power supply method, which involves joint power supply to locomotives from adjacent traction substations, is proposed to enhance power supply capability, regenerative energy utilization, operational safety, and optimize the railway line's longitudinal section layout, also reducing engineering costs [13]. Furthermore, the all-parallel AT power supply method can significantly improve the power supply range and reduce voltage drops in the traction network [14]. Despite extensive studies on the stability of vehicle-grid systems under conventional power supply methods, the all-parallel double-way AT power supply topology still needs to be explored. Given the potential implementation of such a method in railways of high-altitude mountainous areas, investigating its electrical stability is both innovative and necessary.

To address the aforementioned issues, this article introduces a CCM-based impedance modeling method to analyze the harmonic stability of a multiple-vehicle-accessed railway vehiclegrid system. In addition, a multilevel sensitivity function in the frequency domain is presented to identify the most critical element that affects the system stability at the port level, admittance level, and parameters level, respectively. Moreover, a compatibility test procedure for the railway vehicle-grid system is provided, integrating the proposed stability and sensitivity analysis method to support system-level stability analysis and design. The main innovations of this article are summarized as follows.

\begin{enumerate}
  \item We propose the impedance model of a multivehicle connected railway vehicle-grid system under the all-parallel double-way AT feeding power supply topology. The proposed model has been verified to accurately address both LFO and high-frequency HIS issues in the railway system.
  \item Considering the mobility and diverse operating conditions of railway vehicles, we introduced a dynamic sensitivity analysis method. This proposed method can identify the root causes of instability in the system from various levels.
  \item By integrating stability and sensitivity analysis methods, we introduced a compatibility testing process for the railway vehicle-grid system, leading to a system-level stability analysis and parameter design. The proposed method can be used for the compatibility analysis process outlined in the existing EN50388 standard.\\
The rest of this article is organized as follows. Section II presents the CCM-based modeling and stability analysis method of railway vehicle-grid system. Section III presents the
\end{enumerate}

\begin{figure}[H]
\begin{center}
  \includegraphics[alt={},max width=\columnwidth]{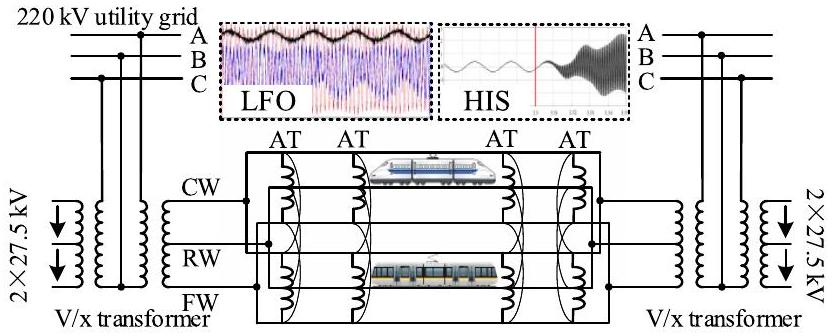}
\caption{Overview the railway vehicle-grid system under all parallel double-way AT feeding power supply.}
\end{center}
\end{figure}

multilevel sensitivity analysis method, and the compatibility test procedure for the system. Case studies are provided and analyzed in Section IV. Finally, Section V concludes this article.

\section*{II. Railway Vehicle-Grid System Modeling and Stability Analysis}
\section*{A. Overview of Railway Vehicle-Grid System}
The main components and structure of the vehicle-grid system under all-parallel AT double-way feeding mode are shown in Fig. 1 [15]. The traction substation is connected to the external 220 kV utility grid, and a three-phase Vx transformer is utilized to step down the voltage to $2 \times 27.5 \mathrm{kV}$. The all-parallel AT traction network comprises a traction substation, traction network, AT transformers, CW, RW, and FW [16]. An AT transformer is usually installed every 20 km along the track. Moreover, the railway system consists of both upward and downward lines. In the full parallel bilateral power supply mode, the AT transformers of the upward and downward are connected in parallel to realize the full parallel power supply. At the same time, two adjacent traction substations are connected with the same phase voltage from the utility grid to achieve bilateral power supply.

\section*{B. All Parallel Double-Way AT Feeding Power Supply System Model}
The traction network is a complex multiconductor chain loop, which is hard to build a complete traction network modeling. However, when the mutual inductance of different conductor lines is ignored, the AT transformer can be equivalent to a node [17], and then a mesh traction network system can be obtained. In addition, in the all paralleled AT power supply mode, the transformer nodes of the upward and downward lines are connected in parallel, so the node potential of paralleled AT transformer is equal. Fig. 2 shows the meshed equivalent circuit of the all parallel AT-powered railway vehicle-grid system.

Since the vehicle is moving along the track, the vehicle is treated as a load or power source at node $k$. The change of its position causes the node admittance matrix of the system to change. Note that the node admittance matrix needs to be generated at different times. To illustrate this further, the equivalent circuit of the system at two different time point $t$ and $t+1$ is illustrated in Fig. 2, where the subscript ga(b) represents the equivalent voltage on the secondary side of the traction substation $\mathrm{A}(\mathrm{B})$, the subscript $t$ represents the system at time $t$, and $Z_{\mathrm{g}}$ denotes the equivalent impedance of the traction substation. Each vehicle,

\begin{figure}[H]
\begin{center}
  \includegraphics[alt={},max width=\columnwidth]{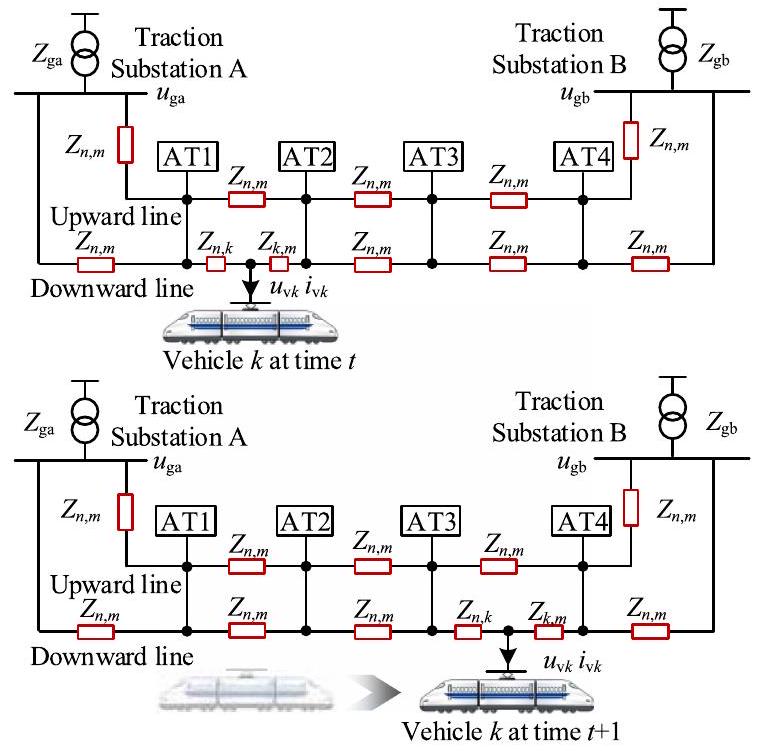}
\caption{Equivalent circuit for railway vehicle-grid interface under the all parallel double-way AT feeding power supply, captured at distinct temporal intervals.}
\end{center}
\end{figure}

represented by $\mathrm{v}_{k}(k=1,2, \ldots, n)$ is connected to the network at a specific node denoted by $k$. The impedance between two AT transformers is defined as $Z_{\mathrm{n}, \mathrm{m}}$. When the vehicle is in the interval of the two AT stations, the impedance of the vehicle node from the left AT is equivalent to $Z_{\mathrm{n}, k}$, and the impedance of the vehicle node from the right AT is equivalent to $Z_{k, \mathrm{~m}}$. These impedances are related to the vehicle's position $l$ in the AT interval and are expressed as

\begin{align*}
Z_{\mathrm{n}, k} & =\frac{l}{L}\left(r_{\mathrm{u}}+j \cdot x_{\mathrm{u}}\right)  \tag{1}\\
Z_{k, \mathrm{~m}} & =\left(1-\frac{l}{L}\right)\left(r_{\mathrm{u}}+j \cdot x_{\mathrm{u}}\right) \tag{2}
\end{align*}

where $L$ is the overall length of an AT segment, $z_{\mathrm{u}}=r_{\mathrm{u}}+j x_{\mathrm{u}}$ represents the unit length impedance of the equivalent circuit of the traction network.

To model the vehicle-grid system, the system is split into the vehicle subsystem and traction network subsystem. The node voltage of vehicle subsystem is represented as $\boldsymbol{u}_{\mathrm{as}}=\left[u_{\mathrm{v} 1}, u_{\mathrm{v} 2}, \ldots\right.$, $\left.u_{\mathrm{v} n}\right]^{\mathrm{T}}$, and node current as $\boldsymbol{i}_{\text{as }}=\left[i_{\mathrm{v} 1}, i_{\mathrm{v} 2}, \ldots, i_{\mathrm{v} n}\right]^{\mathrm{T}}$. The node voltage of traction subsystem is represented as $\boldsymbol{u}_{\mathrm{ps}}=\left[u_{\mathrm{AT} 1}\right.$, $\left.u_{\text{AT2 }}, u_{\text{AT3 }}, u_{\text{AT4 }}, u_{\text{ga }}, u_{\text{gb }}\right]^{\mathrm{T}}$, and the corresponding currents are $\boldsymbol{i}_{\mathrm{ps}}=\left[i_{\mathrm{AT} 1}, i_{\mathrm{AT} 2}, i_{\mathrm{AT} 3}, i_{\mathrm{AT} 4}, i_{\mathrm{ga}}, i_{\mathrm{gb}}\right]^{\mathrm{T}}$. According to the defined nodes, the node admittance matrix $\boldsymbol{Y}_{\text{sys }}$ of the vehicle-grid system can be constructed

\[
\left[\begin{array}{c}
\boldsymbol{i}_{\mathrm{as}}  \tag{3}\\
\boldsymbol{i}_{\mathrm{ps}}
\end{array}\right]=\boldsymbol{Y}_{\mathrm{sys}}\left[\begin{array}{c}
\boldsymbol{u}_{\mathrm{as}} \\
\boldsymbol{u}_{\mathrm{ps}}
\end{array}\right] .
\]

Then, using the Kron reduction [18], the node admittance matrix $\boldsymbol{Y}_{\text{sys }}$ was adopted to eliminate nodes without connected vehicles. Then, (3) can be expressed as

\[
\left[\begin{array}{l}
\boldsymbol{i}_{\mathrm{as}}  \tag{4}\\
\boldsymbol{i}_{\mathrm{ps}}
\end{array}\right]=\left[\begin{array}{ll}
\boldsymbol{Y}_{\mathrm{mm}} & \boldsymbol{Y}_{\mathrm{mn}} \\
\boldsymbol{Y}_{\mathrm{nm}} & \boldsymbol{Y}_{\mathrm{nn}}
\end{array}\right]\left[\begin{array}{l}
\boldsymbol{u}_{\mathrm{as}} \\
\boldsymbol{u}_{\mathrm{ps}}
\end{array}\right] .
\]

Eliminating $\boldsymbol{u}_{\mathrm{ps}}$ from the above matrix, the admittance matrix between $\boldsymbol{U}_{\text{as }}$ and $\boldsymbol{i}_{\text{as }}$ can be derived

\begin{equation*}
\boldsymbol{i}_{\mathrm{as}}=\underbrace{\left(\boldsymbol{Y}_{\mathrm{mm}}-\boldsymbol{Y}_{\mathrm{mn}} \boldsymbol{Y}_{\mathrm{nn}}^{-1} \boldsymbol{Y}_{\mathrm{nm}}\right)}_{\boldsymbol{Y}_{\mathrm{ps}}} \boldsymbol{u}_{\mathrm{as}} \tag{5}
\end{equation*}

\section*{C. Vehicle Subsystem Model}
As vehicles in the railway system are mobile loads, it is considered a typical time-varying system. Because the position, power and operation condition of vehicle will change at different times, it is necessary to divide the vehicle-grid system model into multiple times, and evaluate the stability of the vehicle-grid system at each time in turn.

On the other hand, due to the intellectual property protection of vehicles, generally only the vehicle manufacturer can access to the detailed parameters and circuit topology inside the vehicle. This makes it difficult to obtain the vehicle impedance model by means of impedance modeling. However, vehicle manufacturers can replace the impedance modeling process by providing vehicle impedance measurements or providing data-driven models. Specifically, vehicle manufacturers should provide vehicle impedance measurements results at different operating points, including different traction networks and vehicle operating power. In the stability analysis of practical engineering applications, the vehicle manufacturer should provide these impedance measurement data to the railway operation department to conduct stability analysis by building the vehicle subsystem model.

To construct the vehicle subsystem model, there are some key steps that need to be taken.

\begin{enumerate}
  \item Step 1-Define Global Reference Frame: The first step is to select a node which all impedance matrices shall be referred to. Any node can be selected, and the choice will not affect the stability analysis. If the selected stability analysis is based on source and load impedance equivalents, a logical choice of global reference frame is the source/load interface point. The phase angle at the global reference frame node is defined as $\theta=0$. In this article, the selected global reference frame is the PCC point, which is also the low-voltage side of the traction substation.
  \item Step 2-Traction Power Supply System Power Flow Analysis: First, based on the train schedules, determine the position and operating power of all vehicles in the traction network system at a specific moment. Next, running a power flow calculation will provide the steady-state operation point of the system.
  \item Step 3-Align Vehicle Admittance Models With Global Reference Frame: According to [19], for systems that do not meet the criteria for MFD, it is essential to transform their DQ admittance model to global reference frames. Given the presence of a phase lock loop and dc-link voltage controller within the converter, the DQ impedance model of traction converter does not satisfy the MFD. Hence, its admittance model will depend on the reference frame. The required alignment from the local to global reference frame is provided by the subsequent relations
\end{enumerate}

\begin{equation*}
\boldsymbol{Y}_{\mathrm{v} i d q}^{\mathrm{g}}=T_{d q}^{-1} \boldsymbol{Y}_{\mathrm{v} i d q}^{l} T_{d q} \tag{6}
\end{equation*}

\begin{figure}[H]
\begin{center}
  \includegraphics[alt={},max width=\columnwidth]{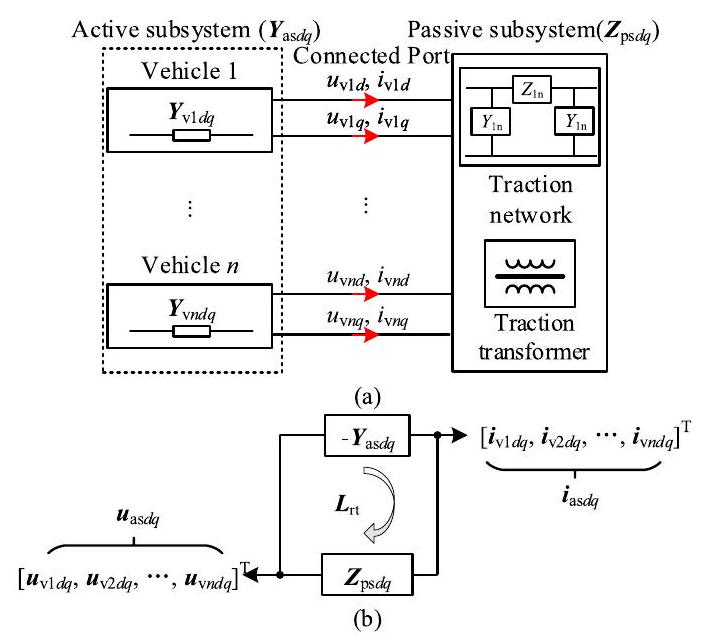}
\caption{Diagram of the railway vehicle-grid system based on the CCM method. (a) Electrical connection. (b) Equivalent MIMO feedback system diagram.}
\end{center}
\end{figure}

\[
T_{d q}=\left[\begin{array}{cc}
\cos \theta_{i} & \sin \theta_{i}  \tag{7}\\
-\sin \theta_{i} & \cos \theta_{i}
\end{array}\right]
\]

where the subscript vidq represents the $d q$ domain model of the $i$ th vehicle, the superscript g stands for the global reference frame, and $l$ denotes the local reference frame. The angle $\theta_{i}$ is the steady-state fundamental voltage angle between component $i$ local reference frame and the global reference frame. In this article, the phase of the PCC node voltage is chosen as the global reference frame and is defined as $\theta=0$.\\
4) Step 4-Build the Admittance Matrix of Vehicle Subsystem $\boldsymbol{Y}_{\text{as }}$ : Based on the impedance results of vehicles in the global reference frame, the active subsystem matrix of vehicles can be obtained

\[
\underbrace{\left[\begin{array}{c}
\boldsymbol{i}_{\mathrm{v} 1 d q}  \tag{8}\\
\boldsymbol{i}_{\mathrm{v} 2 d q} \\
\vdots \\
\boldsymbol{i}_{\mathrm{v} n d q}
\end{array}\right]}_{\boldsymbol{i}_{\mathrm{as} d q}}=\underbrace{\left[\begin{array}{cccc}
\boldsymbol{Y}_{\mathrm{v} 1 d q}^{\mathrm{g}} & & & \\
& \boldsymbol{Y}_{\mathrm{v} 2 d q}^{\mathrm{g}} & & \\
& & \ddots & \\
& & & \boldsymbol{Y}_{\mathrm{v} n d q}^{\mathrm{g}}
\end{array}\right]}_{Y_{\mathrm{as} d q}} \underbrace{\left[\begin{array}{c}
\boldsymbol{u}_{\mathrm{v} 1 d q} \\
\boldsymbol{u}_{\mathrm{v} 2 d q} \\
\vdots \\
\boldsymbol{u}_{\mathrm{v} n d q}
\end{array}\right]}_{\boldsymbol{u}_{\mathrm{as} d q}} .
\]

\section*{D. Frequency-Domain Stability Analysis}
The DQ impedance model of both the traction network and vehicle subsystems are of $2 n$ order, where $n$ depends on the number of nodes in the vehicle access system. The interconnected subsystems in a multivariate feedback system are represented in Fig. 3(a). The elements in both the active and passive subsystems can be obtained using the frequency scan method [20]. To represent the interaction between the two subsystems, the output of $\boldsymbol{Y}_{\text{as } d q}$ should be used as an input to $\boldsymbol{Z}_{\text{ps } d q}$ and vice versa.

The two interconnected subsystems can be further represented as a MIMO negative feedback system as depicted in Fig. 3(b). To evaluate the stability and performance of the MIMO system, the GNSC is employed. The GNSC utilizes the eigenvalues of the system transfer function matrix to assess stability and performance. By analyzing the positions of the eigenvalues, it is possible to determine the system stability and frequency response.

The vehicle-grid system return ratio matrix can be represented as

\begin{equation*}
\boldsymbol{L}_{\mathrm{rt}}=\boldsymbol{Z}_{\mathrm{ps} d q} \boldsymbol{Y}_{\text{as } d q} \tag{9}
\end{equation*}

where $\boldsymbol{L}_{\mathrm{rt}}$ denotes the impedance-return ratio matrix of the system.

For the MIMO feedback system, system stability can be determined by analyzing the characteristic root trajectory of $\boldsymbol{L}_{\mathrm{rt}}$. The vehicle-grid system is considered stable when the phases of all eigenvalues of $\boldsymbol{L}_{\mathrm{rt}}$ intersect with $-180 \times(2 k+1)(k \in \mathrm{Z})$ and have magnitudes less than 0 dB .

\section*{III. Railway Vehicle-Grid System Sensitivity Analysis and Compatibility Test}
In the railway vehicle-grid system, the stability of the system can be affected by various parameters, such as the types of vehicles being operated, the operating modes of the vehicles, and the nominal capacity of the traction transformer, among others. Therefore, to identify the specific vehicle that contributes to system instability and to find effective solutions to address the instability, further analysis of parameter sensitivity is conducted. Based on the proposed stability analysis and sensitivity analysis methods for the vehicle-grid system, a process for determining the compatibility of the vehicle-grid system is presented in this section.

\section*{A. Multilevel Frequency-Domain Sensitivity Analysis}
The multilevel frequency-domain sensitivity analysis method, along with the sensitivity function of $\Lambda_{k}$, can be used to identify the critical vehicle, impedance, and/or parameters of system stability.

By applying eigenvalue decomposition to the return ratio matrix $\boldsymbol{L}_{\mathrm{rt}}$, the following expression is obtained:

\begin{equation*}
\boldsymbol{\Lambda}=\mathbf{W} \boldsymbol{L}_{\mathrm{rt}} \mathbf{V} \tag{10}
\end{equation*}

where $\Lambda$ is the eigenvalue matrix, given by $\Lambda=\operatorname{diag}\left[\Lambda_{1}, \Lambda_{2}, \Lambda_{3}\right.$, $\left.\ldots, \Lambda_{j}\right] ; \mathbf{W}$ is the left eigenvector matrix, with the row representing the unitized left eigenvector of the $j$ th eigenloci; $\mathbf{V}$ is the right eigenvector matrix, with the $j$ th column representing the unitized right eigenvector of $j$ th eigenloci. These matrices are denoted as $\mathbf{W}=\left[\mathrm{W}_{1}, \mathrm{~W}_{2}, \mathrm{~W}_{3}, \ldots, \mathrm{~W}_{j}\right]^{T}$ and $\mathbf{V}=\left[\mathrm{V}_{1}, \mathrm{~V}_{2}, \mathrm{~V}_{3}, \ldots, \mathrm{~V}_{j}\right]$. According to (10), the following equation can be obtained:

\begin{align*}
& \boldsymbol{L}_{\mathrm{rt}} \mathrm{~V}_{k}=\Lambda_{k} \mathrm{~V}_{k}  \tag{11}\\
& \mathrm{~W}_{k} \cdot \mathrm{~V}_{k}=1 \tag{12}
\end{align*}

The $k$ th eigenloci can be expressed as

\begin{equation*}
\Lambda_{k}=\mathrm{W}_{k} \boldsymbol{L}_{\mathrm{rt}} \mathrm{~V}_{k} . \tag{13}
\end{equation*}

\begin{enumerate}
  \item Port-Level Sensitivity Analysis: The port-level participation function of $\Lambda_{k}$ represents the sensitivity of $k$ th about the element $l_{k, j}$ in $\boldsymbol{L}_{\mathrm{rt}}$, which is defined as
\end{enumerate}

$$
\begin{aligned}
p_{k j} & =\frac{\partial \Lambda_{k}}{\partial l_{k, j}} \\
& =\frac{\partial\left(\mathrm{W}_{k} \boldsymbol{L}_{\mathrm{rt}} \mathrm{~V}_{k}\right)}{\partial l_{k, j}}
\end{aligned}
$$

$$
\begin{aligned}
& =\frac{\partial \mathrm{W}_{k}}{\partial l_{k, j}} \boldsymbol{L}_{\mathrm{rt}} \mathrm{~V}_{k}+\mathrm{W}_{k} \frac{\partial \boldsymbol{L}_{\mathrm{rt}}}{\partial l_{k, j}} \mathrm{~V}_{k}+\mathrm{W}_{k} \boldsymbol{L}_{\mathrm{rt}} \frac{\partial \mathrm{~V}_{k}}{\partial l_{k, j}} \\
& =\frac{\partial \mathrm{W}_{k}}{\partial l_{k, j}} \Lambda_{k} \mathrm{~V}_{k}+\mathrm{W}_{k} \frac{\partial \boldsymbol{L}_{\mathrm{rt}}}{\partial l_{k, j}} \mathrm{~V}_{k}+\Lambda_{k} \mathrm{~W}_{k} \frac{\partial \mathrm{~V}_{k}}{\partial l_{k, j}} \\
& =\mathrm{W}_{k} \frac{\partial \boldsymbol{L}_{\mathrm{rt}}}{\partial l_{k, j}} \mathrm{~V}_{k}+\Lambda_{k} \frac{\partial\left(\mathrm{~W}_{k} \mathrm{~V}_{k}\right)}{\partial l_{k, j}} \\
& =\mathrm{W}_{k} \frac{\partial \boldsymbol{L}_{\mathrm{rt}}}{\partial l_{k, j}} \mathrm{~V}_{k} \\
& =w_{k}(j) v_{k}(j)
\end{aligned}
$$

where $j$ represents the $j$ th port, $k$ represents the $k$ th eigenlocus, and $p_{k j}$ denotes the participation of the $j$ th port current in the $k$ th eigenlocus. The level of participation and sensitivity at each port in the system to the eigenlocus can be recognized by the $p_{k j}$. As the apparatus in the active/ passive subsystem is constructed in the $d q$-frame, two ports can be combined by integrating their port-level participation functions to determine the participation of an apparatus [11]. The participation function of each apparatus is represented as follows:

\begin{equation*}
S_{m}^{\Lambda_{k}} \triangleq \sum_{j \in m} p_{k j} \tag{14}
\end{equation*}

where $m$ represents the $m$ th apparatus in the vehicle-grid system. It should be noticed that the frequency-domain sensitivity analysis cannot be simply analyzed at the magnitude or phase crossing frequency of the eigenvalues of $\boldsymbol{L}_{\mathrm{rt}}$ [11]. The key aspect of stability analysis lies in the magnitude of the eigenlocus. Therefore, the critical frequency range is better identified where $\left|\Lambda_{k}\right|>0 \mathrm{~dB}$, rather than just around the phase crossover frequency. Consequently, analyzing sensitivity results exclusively at the crossing frequency is inadequate. In line with this finding, our port-level sensitivity analysis concentrates on the frequency band where the magnitude of critical eigenloci exceeds 0 dB , encompassing the crossing frequency.\\
2) Admittance-Level Sensitivity Analysis: Each vehicle in the vehicle-grid system consists of four admittance/impedance components in the $d q$-frame: $Y_{d d}, Y_{d q}, Y_{q d}$, and $Y_{q q}$. Since vehicles are typically controlled asymmetrically, the effects of these four impedances are different and have various effects on system stability. To further investigate, which admittance component has a significant impact on system stability, an admittance-level sensitivity analysis is conducted. In addition, by applying the shunt or series impedance shaping method to the most critical admittance, the source of instability can be eliminated.

The admittance-level sensitivity analysis involves deriving the sensitivity function of $\Lambda_{k}$ concerning the port device impedance $\boldsymbol{Y}_{\text{as }}(i, j)$

$$
\begin{aligned}
S_{m, d d}^{\Lambda_{k}} & =\frac{\partial \Lambda_{k}}{\partial \boldsymbol{Y}_{\mathrm{as}}(i, i)} \\
& =\sum_{j=1}^{2 m}\left(\frac{\partial \Lambda_{k}}{\partial l_{i, j}} \cdot \frac{\partial l_{i, j}}{\partial \boldsymbol{Y}_{\mathrm{as}}(i, i)}\right)
\end{aligned}
$$

\begin{align*}
& =\sum_{j=1}^{2 m}\left(w_{k}(i) v_{k}(j) \cdot \boldsymbol{Z}_{\mathrm{ps}}(i, j)\right)  \tag{15}\\
S_{m, d q}^{\Lambda_{k}} & =\frac{\partial \Lambda_{k}}{\partial \boldsymbol{Y}_{\mathrm{as}}(i, i+1)} \\
& =\sum_{j=1}^{2 m}\left(\frac{\partial \Lambda_{k}}{\partial l_{i, j}} \cdot \frac{\partial l_{i, j}}{\partial \boldsymbol{Y}_{\mathrm{as}}(i, i+1)}\right) \\
& =\sum_{j=1}^{2 m}\left(w_{k}(i) v_{k}(j) \cdot \boldsymbol{Z}_{\mathrm{ps}}(i+1, j)\right)  \tag{16}\\
S_{m, q d}^{\Lambda_{k}} & =\frac{\partial \Lambda_{k}}{\partial \boldsymbol{Y}_{\mathrm{as}}(i+1, i)} \\
& =\sum_{j=1}^{2 m}\left(\frac{\partial \Lambda_{k}}{\partial l_{i+1, j}} \cdot \frac{\partial l_{i+1, j}}{\partial \boldsymbol{Y}_{\mathrm{as}}(i+1, i)}\right) \\
& =\sum_{j=1}^{2 m}\left(w_{k}(i+1) \cdot v_{k}(j) \cdot \boldsymbol{Z}_{\mathrm{ps}}(i, j)\right)  \tag{17}\\
S_{m, q q}^{\Lambda_{k}} & =\frac{\partial \Lambda_{k}}{\partial \boldsymbol{Y}_{\mathrm{as}}(i+1, i+1)} \\
& =\sum_{j=1}^{2 m}\left(\frac{\partial \Lambda_{k}}{\partial l_{i+1, j}} \cdot \frac{\partial l_{i+1, j}}{\partial \boldsymbol{Y}_{\mathrm{as}}(i+1, i+1)}\right) \\
& =\sum_{j=1}^{2 m}\left(w_{k}(i+1) \cdot v_{k}(j) \cdot \boldsymbol{Z}_{\mathrm{ps}}(i+1, j)\right) \tag{18}
\end{align*}

where the subscripts " $d d, d q, d d$, and $d d$ " denote the subimpedance of the DQ impedance matrix. $\boldsymbol{Y}_{\text{as }}(i, j)$ refers to the element at the $i$ th row and the $j$ th column of the matrix $\boldsymbol{Y}_{\text{as }}$, where $i=2 m-1$.\\
3) Parameter-Level Sensitivity Analysis: Based on the obtained admittance-level sensitivity analysis results, the railway infrastructure manager shares the derived admittance sensitivity functions with the vehicle equipment manufacturers. As the manufacturers have access to all the internal parameters of the vehicle, they can further perform parameter-level sensitivity analysis. Since the units of electrical parameters inside the vehicle are different from those of control parameters, in order to reduce the impact of the units or scales of parameters on the analysis results, different parameters need to be standardized before parameter-level sensitivity analysis. The min-max scaling is used to normalize the different parameters, which scales the data between 0 to 1 range

\begin{equation*}
k_{\mathrm{pnorm}}=\frac{k_{\mathrm{p}}-k_{\mathrm{pmin}}}{k_{\mathrm{pmax}}-k_{\mathrm{pmin}}} \tag{19}
\end{equation*}

where $k_{\text{pnorm }}$ is the normalized value, $k_{\mathrm{p}}$ is the original value, and $k_{\text{pmin }}$ and $k_{\text{pmax }}$ are the minimum and maximum values of the parameter, respectively.

By employing the chain rule, parameter-level sensitivity analysis can be conducted. The ranking of the parameter-level sensitivity analysis results can identify the most critical parameter\\
for system stability. The parameter-level sensitivity is calculated as follows:

\begin{align*}
S_{\mathrm{p}}^{\Lambda_{k}} & =\frac{\partial \Lambda_{k}}{\partial \mathrm{p}} \\
& =\frac{\partial \Lambda_{k}}{\partial \boldsymbol{Y}_{\mathrm{as}}(i, j)} \cdot \frac{\partial \boldsymbol{Y}_{\mathrm{as}}(i, j)}{\partial \mathrm{p}} . \tag{20}
\end{align*}

The derived parameter-level sensitivity analysis results can then be used to optimize the design of controller and circuit parameters.

\section*{B. Procedure for System-Level Compatibility Test for Railway Vehicle-Grid System}
The sensitivity function can be transferred from the port level to the parameter level according to the multilevel sensitivity analysis approach, which can identify the critical parameter of system stability. In addition, sensitivity analysis can be utilized by the railway infrastructure manager and rolling stock suppliers during the compatibility testing process. Fig. 4 provides a flowchart of the system-level compatibility test for the railway vehicle-grid system, which involves the following three primary processes.

\begin{enumerate}
  \item Step 1-System Data Collection: In this process, relevant data regarding the planned railway vehicle-grid system is collected. This includes information about the railway power supply system (traction substation and traction network), vehicle characteristics (such as types and number of vehicles), and system operating conditions (train schedules and feeding arrangements).
\end{enumerate}

Since the vehicle-grid system is time-varying due to the mobility of vehicles, it is transformed into a nontime-varying system by analyzing multiple moments in time. Consequently, the train schedule is divided into multiple time points (a total of $m$ points). At each specified time point, the position, number, operating power, and operation modes of the vehicles within the system are determined. By evaluating the system stability at each time point, an understanding of the system stability throughout the day can be obtained.\\
2) Step 2-System Stability Analysis: In this step, the system stability is analyzed at the specified time point $t_{i}$. The power flow calculation of the railway vehicle-grid system is performed to determine the node voltage and phase of different vehicles [21]. The railway vehicle-grid system is divided into an active subsystem and a passive subsystem by multiple PoCs.

For the passive subsystem, the impedance matrix $\boldsymbol{Z}_{\mathrm{ps}}$ is modeled based on the distances between the vehicles. However, for the active subsystem, the internal circuit structure and control parameters of the vehicles are not accessible due to intellectual property protection. Therefore, it is not possible to obtain the vehicle admittance through modeling. Instead, the vehicle admittance is obtained from impedance data provided by the rolling stock suppliers. These vehicle impedances are then converted to the global $d q$ frame based on the node voltage phase [22]. In addition, the vehicle subsystem admittance matrix $\boldsymbol{Y}_{\text{as }}$ is constructed based on the location of vehicle access.

\begin{figure}[H]
\begin{center}
  \includegraphics[alt={},max width=\columnwidth]{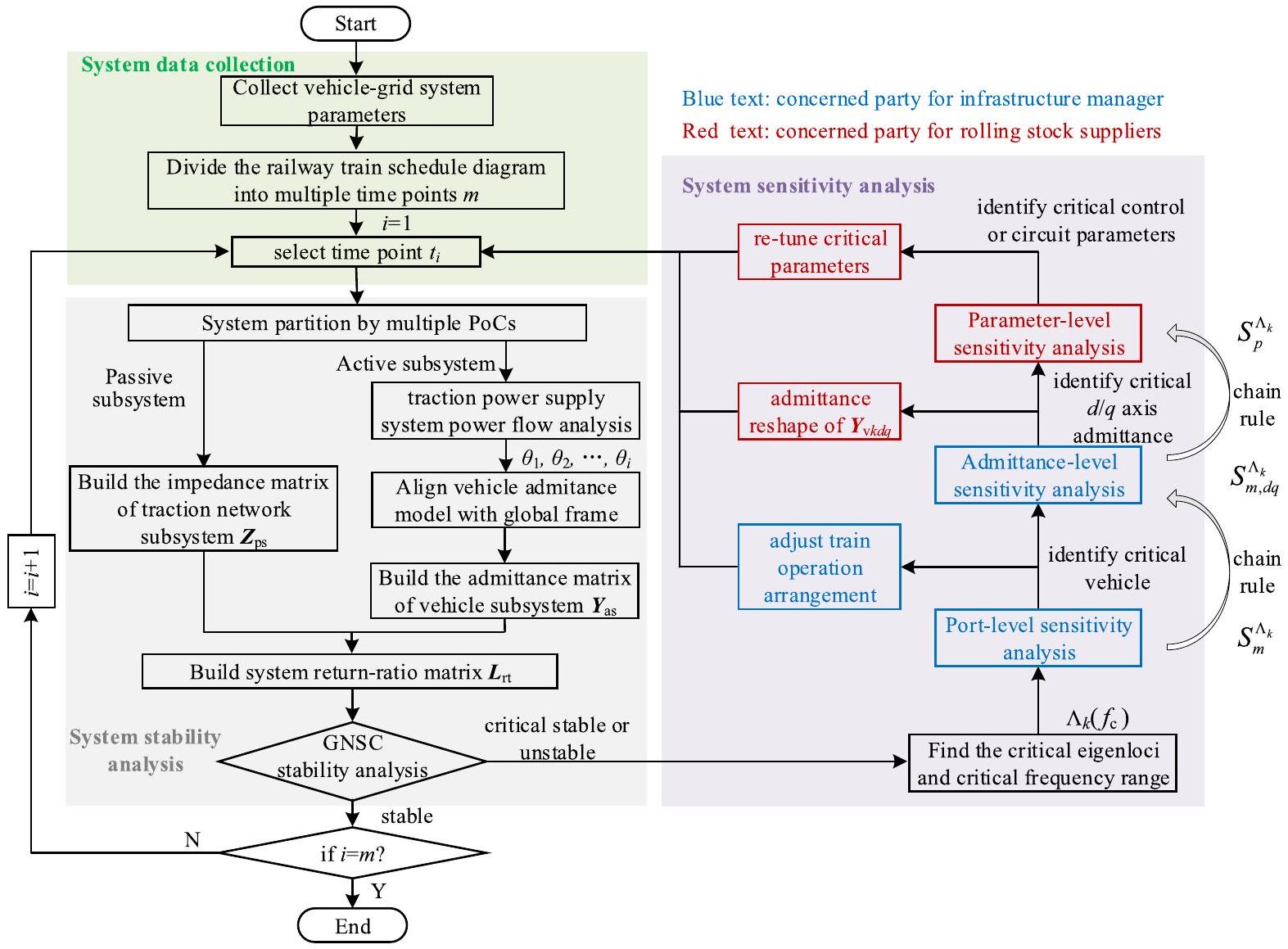}
\caption{Flowchart of system-level compatibility test for railway vehicle-grid system.}
\end{center}
\end{figure}

Then, the system return-ratio matrix, $\boldsymbol{L}_{\mathrm{rt}}$, is constructed from the two subsystems, and system stability is analyzed by the GNSC. If all the characteristic roots of the $\boldsymbol{L}_{\mathrm{rt}}$ satisfy the stability requirement, the stability analysis of the system at this time point is considered complete. The stability of the system at the next time point is then analyzed based on the train schedule. However, if there are unstable or critical stable eigenlocus of $\boldsymbol{L}_{\mathrm{rt}}$, it is necessary to identify the critical eigenloci and their corresponding critical frequency ranges. In addition, the system sensitivity can be analyzed to further understand the impact of different parameters on system stability.\\
3) Step 3-Multilevel Sensitivity Analysis and Stability Enhancement: Port-level sensitivity analysis is initially conducted using (14), allowing the identification of critical vehicles that impact system stability. With this information, the infrastructure manager can optimize system stability by adjusting the operation arrangement of vehicles, such as modifying travel intervals or replacing vehicles with different types. The optimized system is then re-evaluated for stability.

Moreover, for the identified critical vehicles, the infrastructure manager can perform an admittance-level sensitivity analysis to identify how the vehicle is affected by the $d$-axis or $q$-axis impedance. This analysis enables the optimization of the impedance model using techniques, such as impedance reshaping [23], or adding active components in the railway system [24]. For example, Chen and Lin et al., formulated a\\
sequence component-based multiobjective optimization problem for the cascaded energy storage system to maximize the energy efficiency improvement while ensuring that the power quality meets the standard under complex operating conditions. In addition, the infrastructure manager can share the admittancelevel sensitivity function with the rolling stock supplier, enabling parameter-level sensitivity analysis. This analysis helps determine the influence of circuit or control parameters on critical eigenloci.

Lastly, the system stability analysis is iteratively conducted until stability is achieved at the given time point. This iterative process ensures that the system remains stable and performs optimally.

\section*{C. Comparison With the Existing Methods}
While several system-level stability analysis and frequencydomain sensitivity analysis methods have been proposed, these methods vary in terms of algorithm complexity and method applicability. To highlight the innovation in this article, we compare the proposed methods with existing frequency-domain sensitivity analysis methods.

\begin{enumerate}
  \item Compared With the Classical State-Based Sensitivity Analysis Methods: The methods in [25] and [26] identify the impedance model of converters through the vector fitting algorithm and then obtain the system dynamic state-space matrix. By\\
using classical state-based sensitivity analysis on the dynamic state-space matrix, unstable operating conditions and poorly damped oscillatory modes can be identified. Compared with the method proposed in this article, it can be found that
  \item The model order of classical state-based sensitivity analysis is higher.\\
State-based participation analysis analyzes the characteristic roots of the state equation to perform participation factor. The order of the state equation depends on the number of system states. In contrast, for the method proposed in this article, the model order depends on the number of selected ports. Generally, these selected ports are the access points of vehicles in the system, which results in a significantly lower quantity compared with the number of system state variables.
  \item The calculation steps involved in classical state-based sensitivity analysis are more complex.\\
Classical state-based sensitivity analysis requires system identification of the frequency domain transfer function and solving high-order algebraic equations to calculate poles for stability analysis in black box systems. This approach involves pole-zero cancellation phenomena and data fitting challenges. However, the method proposed in this article simplifies the process by directly calculating the impedance return ratio matrix, eliminating the need for system model identification and pole calculation.
  \item Compared With the Residue Representation-Based Impedance Sensitivity Methods: The method proposed in [10] utilizes impedance sensitivity analysis based on residue representations to trace the root causes of oscillation issues in multiple power electronics integrated power systems. Compared with the method proposed in this article, the impedance sensitivity analysis based on residue representations relies on known transfer function expressions or estimated residues, which is inapplicable for the pure black box systems or faces model identification processes.
  \item Compared With the Application of Sensitivity in Impedance Stability Analysis of Converter Systems: Both converter systems and vehicle-grid systems are power-electronic energy conversion systems, and sensitivity analysis is a wellestablished and common analytical method. Therefore, there are many similarities between this manuscript and the application of sensitivity analysis of converter systems. However, due to the uniqueness of the study objects, there are two main differences in application, listed as follows.
  \item Compared to the converter systems, the application of sensitivity analysis in railway vehicle-grid systems requires a smaller frequency-scan step to capture the impedance information of locomotives accurately.\\[0pt]
The railway vehicle-grid systems have more intricate frequency coupling effects than conventional three-phase converter systems. This complexity primarily arises from the single-phase circuit configuration and the distinctive control strategy employed in the PLL. On the one hand, the vehicle grid system is a single-phase system analogous to a three-phase system where only two phases are utilized. This situation is typically considered an unbalanced three-phase or three-phase asymmetric system [27]. On the other hand, in single-phase systems,\\[0pt]
the added quadrature signal generator in single-phase PLL increases the asymmetrical level of the whole system. Therefore, the output of single-phase PLL contains additional harmonic components [28], which is not observed in the three-phase VSC systems.
\end{enumerate}

As a result, the impedance characteristics in the lowfrequency range of railway vehicle-grid systems are more intricate than those in typical three-phase converter systems. This complexity becomes particularly evident when comparing the DQ impedance/admittance characteristics of single-phase and three-phase VSC systems.

Therefore, for railway vehicle-grid systems, minimizing the analyzed frequency step during sensitivity analysis is crucial. This approach ensures that critical frequency-domain information is not overlooked. For instance, in [12], the analyzed frequency step for the three-phase VSC system is set at 12 Hz , carried out from 1 to 50 Hz with five logarithmically spaced points. In contrast, our study adopts a more refined frequency step of 1 Hz in the low-frequency range, spanning from 1 to 50 Hz with 50 logarithmically spaced points. This finer granularity in frequency analysis is essential to accurately capture the complex dynamics of single-phase vehicle-grid systems.

\begin{enumerate}
  \item Compared with the converter systems, the application of sensitivity analysis in railway vehicle-grid systems requires a reduced time step for a more accurate and thorough analysis of system stability.\\
Converter systems are characterized by stationary locations and hourly power fluctuations. For example, the power output of photovoltaic systems varies throughout the day, influenced by sunlight exposure. In contrast, vehicle-grid systems exhibit dynamic changes due to the continuous movement of vehicles. In addition, the operation of the vehicle-grid system frequently alternates among TM (acting as a variable power load), POM (a constant low-power load), and RBM (serving as a power supply). Given the rapid switching between these modes, frequencydomain sensitivity analysis of vehicle-grid systems requires a finer time-step approach to capture these transient dynamics accurately.
\end{enumerate}

Overall, the proposed method in this article demonstrates distinct advantages over existing methods by offering a lower model order, more straightforward calculations, and applicability to the railway vehicle-grid systems.

\section*{IV. Verification}
This section validates the proposed stability and sensitivity analysis method using real-world railway train schedules. HIL experiments are used to analyze the designed cases, considering the high voltage of the traction network and the high-power characteristics of the vehicles. The HIL platform, as depicted in Fig. 5 is utilized for the railway vehicle-grid system. It consists of a host computer, NI-PXIe-FPGA-7868R, NI-PXIe-FPGA7846R, an oscilloscope, and an input/output board [29]. The host computer configures the hardware setup using Starsim software and then transfers it to the FPGA-based real-time simulator. The real-time controller executes the control circuits of the different vehicles and generates the corresponding pulsewidth modulation

\begin{figure}[H]
\begin{center}
  \includegraphics[alt={},max width=\columnwidth]{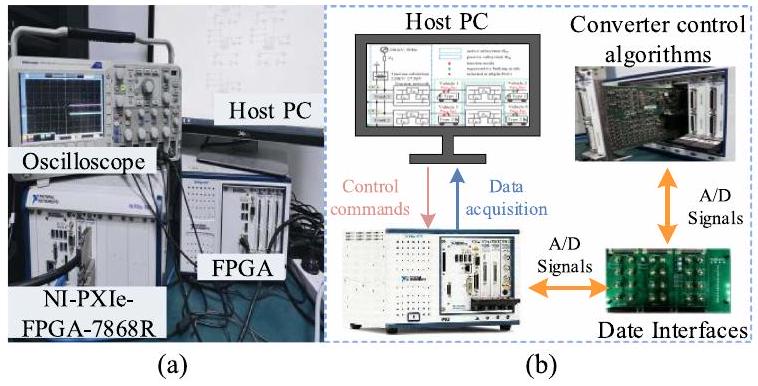}
\caption{HIL platform for railway vehicle-grid system. (a) Platform configuration. (b) Detailed implementation.}
\end{center}
\end{figure}

\begin{table}[!t]
\centering
\caption{Parameters of the Railway Vehicle-Grid System}
\label{tab:railway_vehicle_grid_parameters}
\scriptsize
\setlength{\tabcolsep}{2pt}
\renewcommand{\arraystretch}{1.08}
\begin{tabular}{@{}p{0.23\columnwidth}p{0.48\columnwidth}p{0.23\columnwidth}@{}}
\toprule
Item & Parameters & Values \\
\midrule
\multirow{8}{0.23\columnwidth}{\centering Traction power supply system}
 & Grid voltage & 220 kV \\
 & Grid-side equivalent inductance & 0.7 mH \\
 & Rated power & 40 MVA \\
 & Rated voltage & $220/27.5\,\mathrm{kV}$ \\
 & No-load loss & 22.395 kW \\
 & No-load current & 0.06\% \\
 & Short-circuit impedance & 10.34 \\
 & OCLs & CTMH150 \\
\midrule
\multirow{7}{0.23\columnwidth}{\centering Vehicle}
 & Onboard transformer $(R_{\mathrm{n}}, L_{\mathrm{n}})$ & $(0.14\,\Omega, 5.4\,\mathrm{mH})$ \\
 & DC-side capacitance & $C_{\mathrm{d}}=9\,\mathrm{mF}$ \\
 & Switching frequency & $f_{\mathrm{sw}}=1000\,\mathrm{Hz}$ \\
 & DC-side reference voltage & $u_{\mathrm{dc,ref}}=3600\,\mathrm{V}$ \\
 & PLL bandwidth & $BW_{\mathrm{PLL}}=30\,\mathrm{Hz}$ \\
 & Voltage control loop $(K_{\mathrm{P}u}, K_{\mathrm{I}u})$ & $(0.5\,\mathrm{S}, 5\,\mathrm{S/s})$ \\
 & Current control loop $(K_{\mathrm{P}i}, K_{\mathrm{I}i})$ & $(1.7\,\Omega, 50\,\Omega/\mathrm{s})$ \\
\bottomrule
\end{tabular}
\end{table}

\begin{figure}[H]
\begin{center}
  \includegraphics[alt={},max width=\columnwidth]{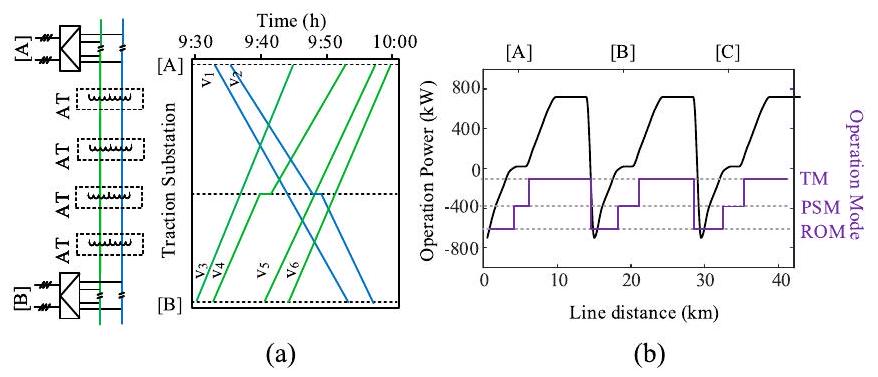}
\caption{Railway vehicle-grid system train schedules. (a) Railway vehicle-grid system train schedules. (b) Vehicle operation mode at different line positions.}
\end{center}
\end{figure}

(PWM) pulses. The HIL experimental results are saved by the host computer and oscilloscope for further analysis. The railway vehicle-grid parameters are provided in Table I, where the main circuit and control parameters of vehicle are selected according to the real parameters of CRH5 EMUs' converters [6], [30].

\section*{A. System Description}
The train schedule from 9:30 to 10:00 is depicted in Fig. 6(a), where the horizontal axis denotes the time in hours, the vertical axis represents the distance of the line, and $[\mathrm{A}],[\mathrm{B}]$, and $[\mathrm{C}]$ correspond to the three traction substations. The green and blue lines denote the vehicles on the upward and downward lines,

\begin{figure}[H]
\begin{center}
  \includegraphics[alt={},max width=\columnwidth]{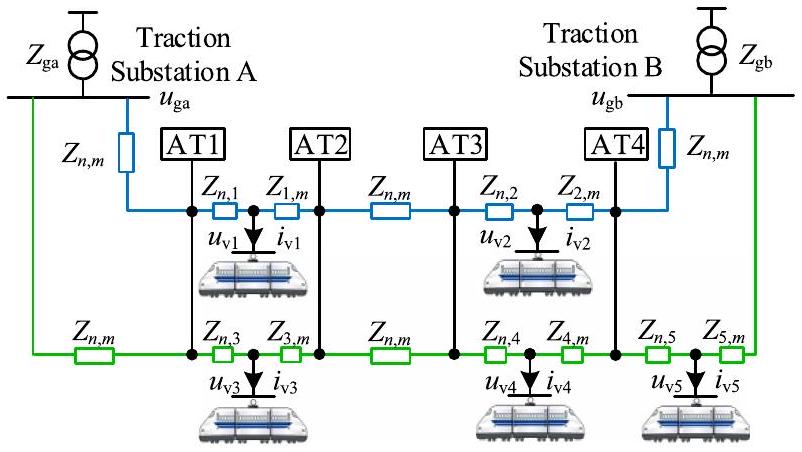}
\caption{Schematic diagram of the railway vehicle-grid system of Case 1.}
\end{center}
\end{figure}

respectively, while the power supply topology of the system is shown on the right-hand side. To facilitate the subsequent stability analysis, the train schedule is divided into 30 time points. Based on the intersection of these time points and the vehicles, the operating power and modes of vehicles can be determined according to Fig. 6(b).

Following the flowchart in Fig. 4, a system-level compatibility test is performed for the vehicle-grid system configuration at 30 time points. Here, the system at time 9:44 and 9:50 are provided as the following analyzed cases.

\begin{enumerate}
  \item Case 1: At 9:44, the vehicle-grid system configuration includes two vehicles ( $\mathrm{v}_{1}$ and $\mathrm{v}_{2}$ ) with power outputs of 560 and 720 kW , respectively, connected to the downward line. In addition, three vehicles $\left(\mathrm{v}_{3}-\mathrm{v}_{5}\right)$ with power outputs of $-100,720$, and 250 kW , respectively, are connected to the upward line.
  \item Case 2: At 9:50, the vehicle-grid system configuration includes two vehicles ( $\mathrm{v}_{1}$ and $\mathrm{v}_{2}$ ) with power outputs of -700 and -720 kW , respectively, connected to the downward line. In addition, three vehicles ( $\mathrm{v}_{4}-\mathrm{v}_{6}$ ) with power outputs of $-270,720$, and 440 kW , respectively, are connected to the upward line. (In order to simulate the phenomenon of high-frequency instability, the equivalent inductance of onboard transformer of the sixth vehicle is set to 0.54 mH .)\\[0pt]
The railway vehicle-grid system in Case 1 is presented in Fig. 7, where the system contains five vehicles, with two vehicles operated on the downward line and three on the upward line. The circuit structure and control block diagram of the traction power supply system in the vehicle is shown in Fig. 8. One vehicle consists of several components, including an onboard traction transformer, traction rectifier, dc-link circuit, traction inverter, and traction motor [31].
\end{enumerate}

\section*{B. System Stability Analysis}
According to Fig. 7, the access location of each vehicle, the AT station, and the exit location of the traction substations are selected as nodes to be analyzed. The node voltage is selected as $\boldsymbol{u}_{d q}=\left[\boldsymbol{u}_{\mathrm{v} 1 d q}, \boldsymbol{u}_{\mathrm{v} 2 d q}, \boldsymbol{u}_{\mathrm{v} 3 d q}, \boldsymbol{u}_{\mathrm{v} 4 d q}, \boldsymbol{u}_{\mathrm{v} 5 d q}, \boldsymbol{u}_{\mathrm{AT} 1 d q}, \boldsymbol{u}_{\mathrm{AT} 2 d q}\right.$, $\boldsymbol{u}_{\text{AT3 } d q}, \boldsymbol{u}_{\text{AT4 } d q}, \boldsymbol{u}_{\text{ga } d q}$, and $\boldsymbol{u}_{\text{gb } d q}$ ]. The node current is selected as $\boldsymbol{i}_{d q}=\left[\boldsymbol{i}_{\mathrm{v} 1 d q}, \boldsymbol{i}_{\mathrm{v} 2 d q}, \boldsymbol{i}_{\mathrm{v} 3 d q}, \boldsymbol{i}_{\mathrm{v} 4 d q}, \boldsymbol{i}_{\mathrm{v} 5 d q}, \boldsymbol{i}_{\mathrm{AT} 1 d q}, \boldsymbol{i}_{\mathrm{AT} 2 d q}, \boldsymbol{i}_{\mathrm{AT} 3 d q}\right.$, $\boldsymbol{i}_{\text{AT } 4 d q}, \boldsymbol{i}_{\text{ga } d q}$, and $\boldsymbol{i}_{\text{gb } d q}$ ].

\begin{figure}[H]
\begin{center}
  \includegraphics[alt={},max width=\columnwidth]{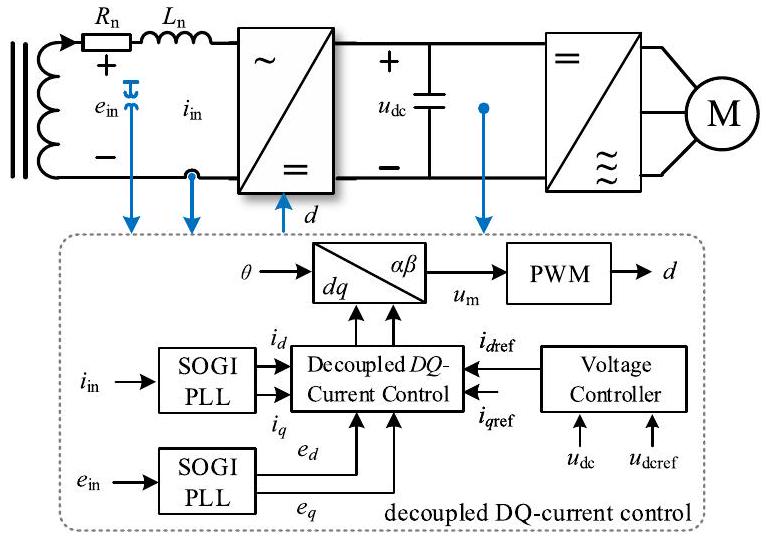}
\caption{Circuit topology and control diagram of vehicle traction power supply system.}
\end{center}
\end{figure}

Then, constructing the node admittance matrix of the system according to the selected node voltage and current relationship

$$
\boldsymbol{i}_{d q}=\boldsymbol{Y}_{\mathrm{sys}} \boldsymbol{u}_{d q}
$$

The nodes can be categorized into two types: those connecting active devices, such as vehicles, and those linking to passive network elements, including AT and traction substation exit locations

$$
\left[\begin{array}{c}
\boldsymbol{i}_{\mathrm{as}} \\
\boldsymbol{i}_{\mathrm{ps}}
\end{array}\right]=\boldsymbol{Y}_{\mathrm{sys}}\left[\begin{array}{c}
\boldsymbol{u}_{\mathrm{as}} \\
\boldsymbol{u}_{\mathrm{ps}}
\end{array}\right] .
$$

To extract the node voltage-current relationship for active devices, the Kron reduction method is employed

$$
\left[\begin{array}{c}
\boldsymbol{i}_{\mathrm{as}} \\
\boldsymbol{i}_{\mathrm{ps}}
\end{array}\right]=\left[\begin{array}{ll}
\boldsymbol{Y}_{\mathrm{mm}} & \boldsymbol{Y}_{\mathrm{mn}} \\
\boldsymbol{Y}_{\mathrm{nm}} & \boldsymbol{Y}_{\mathrm{nn}}
\end{array}\right]\left[\begin{array}{l}
\boldsymbol{u}_{\mathrm{as}} \\
\boldsymbol{u}_{\mathrm{ps}}
\end{array}\right]
$$

The extracted node admittance matrix is transformed to the $d q$ frame to derive $Y_{\mathrm{ps}}$

\begin{equation*}
\boldsymbol{i}_{\mathrm{as}}=\underbrace{\left(\boldsymbol{Y}_{\mathrm{mm}}-\boldsymbol{Y}_{\mathrm{mn}} \boldsymbol{Y}_{\mathrm{nn}}^{-1} \boldsymbol{Y}_{\mathrm{nm}}\right)}_{\boldsymbol{Y}_{\mathrm{ps}}} \boldsymbol{u}_{\mathrm{as}} . \tag{21}
\end{equation*}

The transfer function matrix of active subsystem $Y_{\text{as }}$ can be constructed as follows:

\[
\boldsymbol{Y}_{\text{as }}=\left[\begin{array}{ccccc}
\boldsymbol{Y}_{\mathrm{v} 1 d q}^{\mathrm{g}} & & & &  \tag{22}\\
& \boldsymbol{Y}_{\mathrm{v} 2 d q}^{\mathrm{g}} & & & \\
& & \boldsymbol{Y}_{\mathrm{v} 3 d q}^{\mathrm{g}} & & \\
& & & \boldsymbol{Y}_{\mathrm{v} 4 d q}^{\mathrm{g}} & \\
& & & & \boldsymbol{Y}_{\mathrm{v} 5 d q}^{\mathrm{g}}
\end{array}\right]
\]

where $\boldsymbol{Y}_{\mathrm{v} \text{ idq }}^{\mathrm{g}}$ is the impedance of each vehicle converted to the global $d q$ frame, and nondiagonal elements in $\boldsymbol{Y}_{\text{as }}$ are zero.

The vehicle grid system return ratio matrix can be represented as

\begin{equation*}
\boldsymbol{L}_{\mathrm{rt}}=\boldsymbol{Z}_{\mathrm{ps}} \boldsymbol{Y}_{\mathrm{as}} \tag{23}
\end{equation*}

where $L_{\mathrm{rt}}$ denotes the impedance-return ratio matrix of the system, and $I_{2 n}$ is the $2 n$-order unit matrix.

Then, the return ratio matrix $L_{\mathrm{rt}}$ is established. Using the GNSC, the system stability can be analyzed by evaluating the eigenlocus trajectory of $L_{\mathrm{rt}}$. The stability results for Case 1 are

\begin{figure}[H]
\begin{center}
  \includegraphics[alt={},max width=\columnwidth]{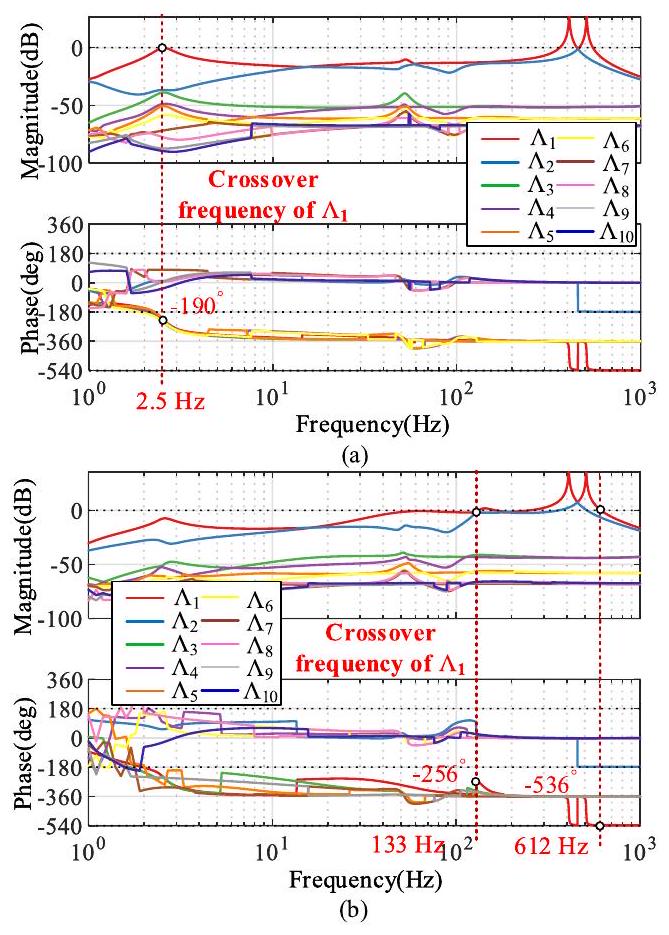}
\caption{Vehicle-grid system stability analysis results. (a) Case 1. (b) Case 2.}
\end{center}
\end{figure}

\begin{figure}[H]
\begin{center}
  \includegraphics[alt={},max width=\columnwidth]{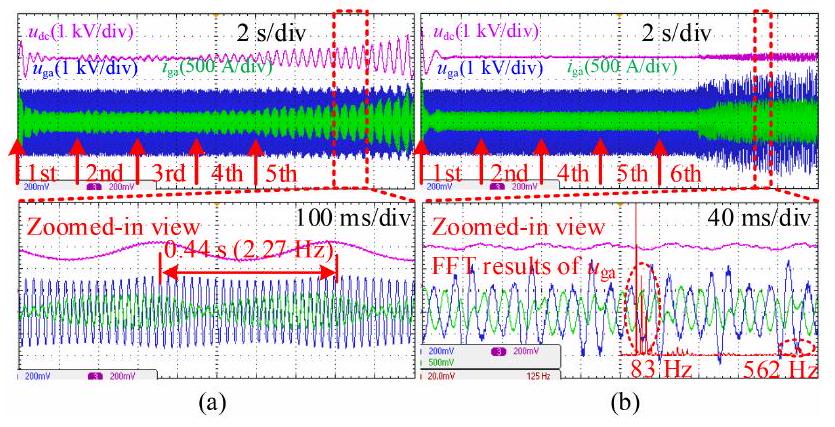}
\caption{Waveforms of line-side voltage and current, and DC-link voltage of vehicle. (a) Case 1. (b) Case 2.}
\end{center}
\end{figure}

shown in Fig. 9. In this case, there are five vehicles, resulting in a tenth-order system model in the $d q$-frame. It can be found that the amplitude of eigenlocus $\Lambda_{1}$ is the largest among ten eigenlocus and has an intersection with 0 dB amplitude at 2.5 Hz , with a phase of $-190^{\circ}$, indicating that the system is unstable at 2.5 Hz .

The stability of Cases 1 and 2 is assessed by sequentially introducing vehicles into the system and observing the waveforms of line-side voltage and current, as well as the dc-side voltage of the vehicles.

The results for Case 1 are shown in Fig. 10(a), where the five vehicles are connected to the system in sequence. It can be observed that voltage and current experience fluctuations when the first three vehicles are connected, but eventually stabilize. However, when the fourth and fifth vehicles are added to the system, continuous oscillations in system voltage and current occur, and the oscillation amplitude gradually increases. By analyzing the oscillation frequency of the dc-side voltage of the vehicles, it is determined to be 2.27 Hz . These results indicate\\
that the system is unstable under Case 1, which aligns with the stability analysis results.

The stability of Case 2 is assessed by sequentially introducing vehicles into the system and observing the waveforms of line-side voltage and current, as well as the dc-side voltage of the vehicles. The results for Case 2 are shown in Fig. 10(b), where the five vehicles are connected to the system in sequence. It can be found that when the first four vehicles are connected to the system, the system can maintain stable operation. However, when the fifth vehicle is connected, the system voltage continues to oscillate, and the frequency of oscillation is mainly 83 and 562 Hz . It is worth noting that the vehicle-network system stability analysis results in Fig. 9(b) are established under the $d q$ frame, while Fig. 10(b) presents the FFT analysis of the ac-side voltage, aligning with the results in $\alpha \beta$ frame. Therefore, there is a 50 Hz frequency difference between the crossover frequency and the ac voltage harmonic frequency.

\section*{C. Port-Level Sensitivity Analysis}
To further investigate the primary contributor to poor damping and identify methods to enhance damping through sensitivity analysis, it is necessary to determine the specific vehicle responsible for system stability. Fig. 11(a) shows the dynamic sensitivity results of vehicles in the system at different times.

Utilizing the port-level sensitivity analysis, the participation functions for ports 1-2, 3-4, 5-6, 7-8, and 9-10 are grouped to represent the participation of vehicles 1-5, respectively. In Case 1 , the participation for various ports regarding $\Lambda_{1}$ is depicted in Fig. 11(b), emphasizing the low-frequency range. With the crossover frequency of $\Lambda_{1}$ around 2 Hz , both vehicles 2 and 4 show pronounced participation in this range, while vehicles 3 and 5 have minimal impact on system stability. Therefore, the admittance-level and parameter-level sensitivity results are analyzed at only the crossing frequency point. Conversely, in Case 2, focusing on the crossover frequency of $\Lambda_{1}$, Fig. 11(c) highlights that vehicle 6 has the most significant participation in the high-frequency range, whereas other vehicles contribute less to system stability.

To evaluate the influence of different vehicles on system stability in Case 1, all vehicles are sequentially removed from the system at time $t$. In Fig. 12(b) and (d), it is observed that the system oscillations disappear and eventually stabilize after removing vehicles 2 and 4. In Fig. 12(a) and (e), after removing vehicles 1 and 5, the amplitude of the system oscillations decreases and eventually stabilizes, but the duration of the oscillations is longer compared with Fig. 12(b) and (d). This indicates that vehicles 2 and 4 have a greater impact on system stability than vehicles 1 and 5 . In Fig. 12(c), the system continues to oscillate after removing vehicle 3 , indicating that vehicle 3 have a lesser impact on system stability. Based on the above analysis, it can be concluded that vehicles 2 and 4 have the greatest influence on system stability, followed by vehicles 1 and 5 , and finally vehicle 3 . This finding is consistent with the results obtained from the port-level sensitivity analysis.

\begin{figure}[H]
\begin{center}
  \includegraphics[alt={},max width=\columnwidth]{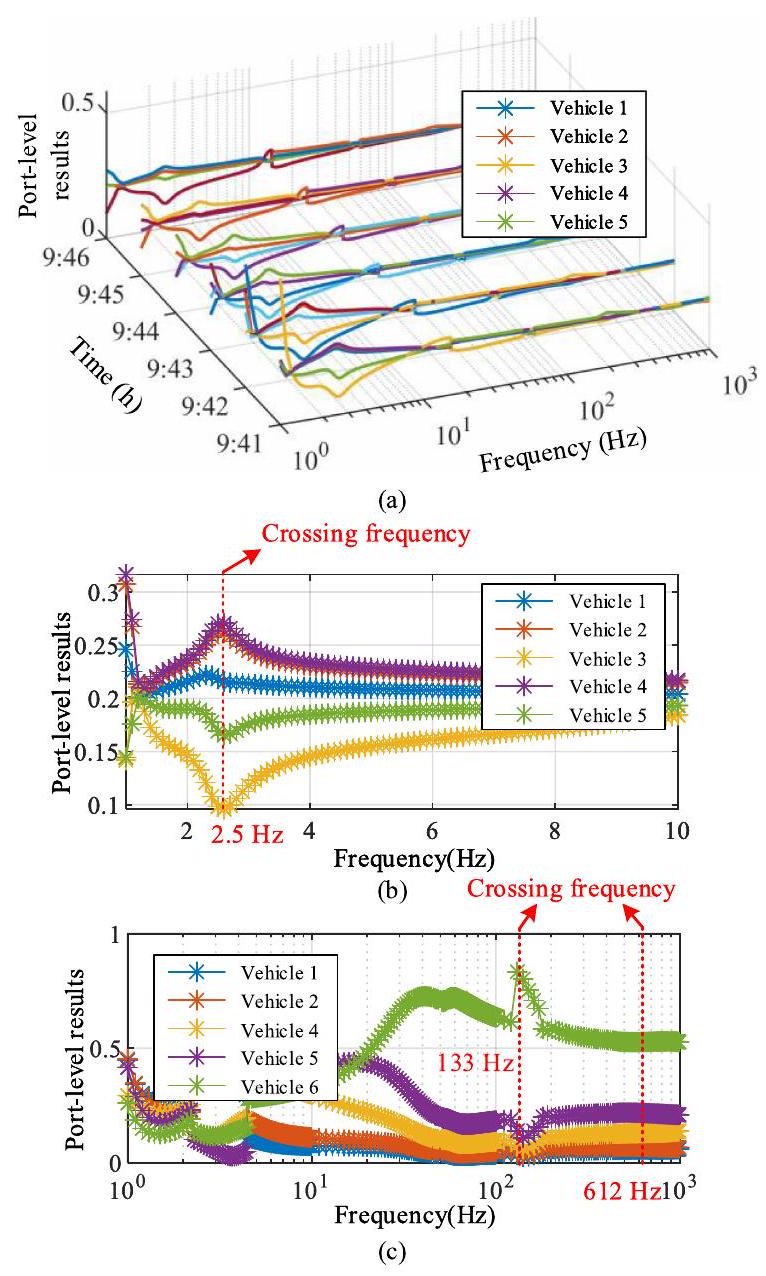}
\caption{Port-level sensitivity analysis results. (a) Port-level sensitivity analysis at time 9:41-9:46. (b) Port-level sensitivity analysis of Case 1. (c) Port-level sensitivity analysis of Case 2.}
\end{center}
\end{figure}

\begin{figure}[H]
\begin{center}
  \includegraphics[alt={},max width=\columnwidth]{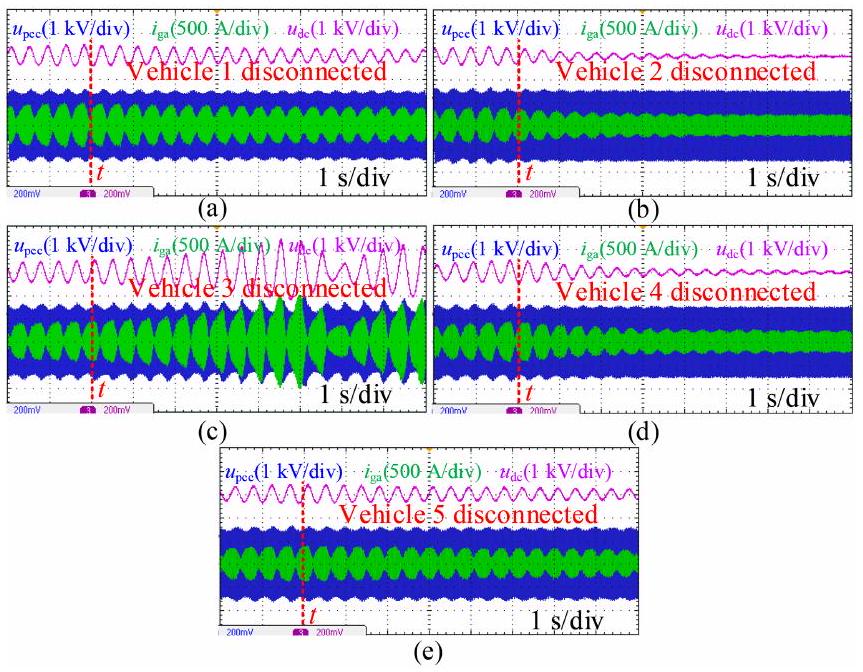}
\caption{Waveforms of line-side voltage and current at PCC, and DC-link voltage of vehicle of Case 1. (a) Disconnect vehicle 1 at $t$. (b) Disconnect vehicle 2 at $t$. (c) Disconnect vehicle 3 at $t$. (d) Disconnect vehicle 4 at $t$. (e) Disconnect vehicle 5 at $t$.}
\end{center}
\end{figure}

\begin{figure}[H]
\begin{center}
  \includegraphics[alt={},max width=\columnwidth]{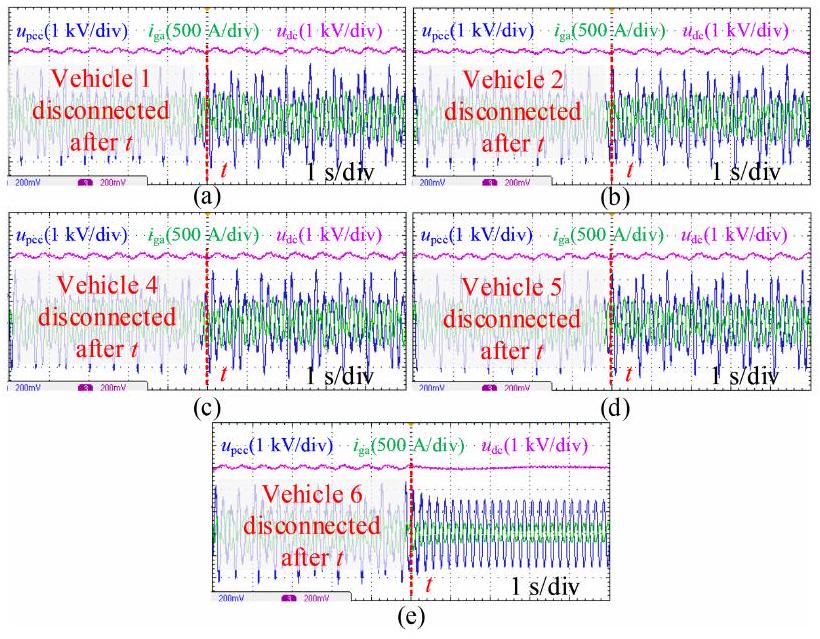}
\caption{Waveforms of line-side voltage and current at PCC, and DC-link voltage of vehicle of Case 2. (a) Disconnect vehicle 1 after $t$. (b) Disconnect vehicle 2 after $t$. (c) Disconnect vehicle 4 after $t$. (d) Disconnect vehicle 5 after $t$. (e) Disconnect vehicle 6 after $t$.}
\end{center}
\end{figure}

To evaluate the influence of different vehicles on system stability in Case 2, all vehicles are sequentially removed from the system after time $t$. It can be seen from Fig. 13(a)-(d) that after high-frequency HIS issue occurs in the system, whether removing vehicle $1,2,4$, or 5 has little impact on system stability. However, in Fig. 13(e), once the vehicle 6 is removed, the high-frequency HIS issue disappears. The results show that vehicles $1-5$ have little influence on the high frequency stability of the system, but vehicle 6 has great influence. The experimental results align with the port-level sensitivity analysis. Moreover, the underlying cause for this phenomenon is that in the design of Case 2, we deliberately reduced the on-board transformer capacity of vehicle 6. This reduction makes the equivalent inductance of vehicle 6 smaller than that of vehicles 1-5, which results in vehicle 6 having a weaker ability to suppress high-frequency harmonics. Consequently, the generated high frequency harmonic components from vehicle 6 are injected into the traction network, leading high-frequency HIS issues occur in the system.

\section*{D. Admittance-Level Sensitivity Analysis}
To investigate the effect of the $d$-axis or $q$-axis impedance on system stability, the admittance-level sensitivity analysis is performed by the infrastructure manager. In Case 1, the impact of the DQ admittance of different vehicles on the eigenlocus $\Lambda_{1}$ at the crossover frequency is depicted in Fig. 14(a), where the admittance amplitude represents its influence on system stability. It can be observed that the $d$-axis or $q$-axis admittance of vehicles 2 and 4 has the greatest impact on system stability, followed by vehicles 1 and 5 . The impact of vehicle 3 is minimal. This finding aligns with the results obtained from the port-level sensitivity analysis. Similarly, as for Case 2, the impact of the DQ admittance of different vehicles on the eigenlocus $\Lambda_{1}$ at the crossover frequency is depicted in Fig. 14(b). It can be observed that the $d$-axis or $q$-axis admittance of vehicle 6 has the greatest impact on system stability.

\begin{figure}[H]
\begin{center}
  \includegraphics[alt={},max width=\columnwidth]{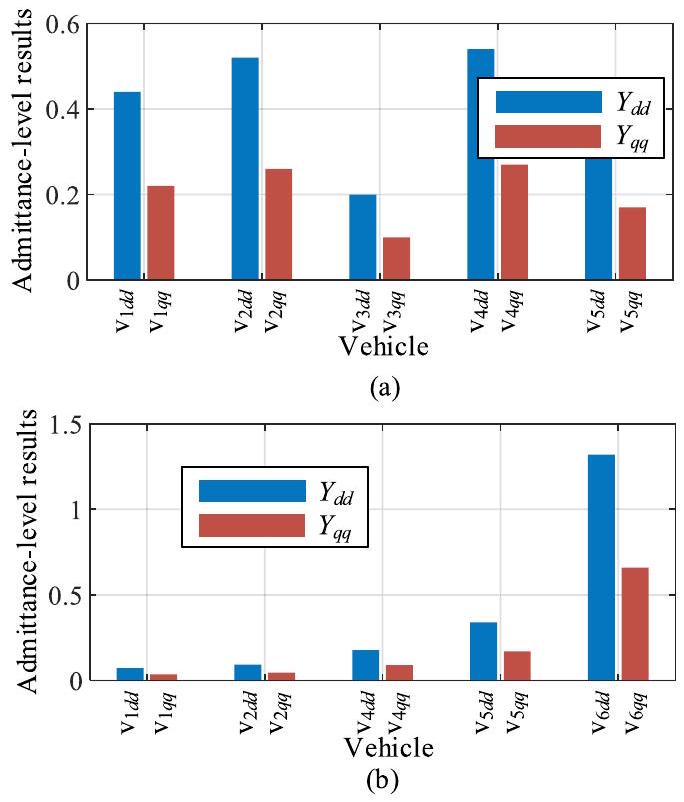}
\caption{Admittance-level sensitivity. (a) Analysis for $\Lambda_{1}$ at 2.5 Hz in Case 1 . (b) Analysis for $\Lambda_{1}$ at 83 Hz in Case 2.}
\end{center}
\end{figure}

\begin{figure}[H]
\begin{center}
  \includegraphics[alt={},max width=\columnwidth]{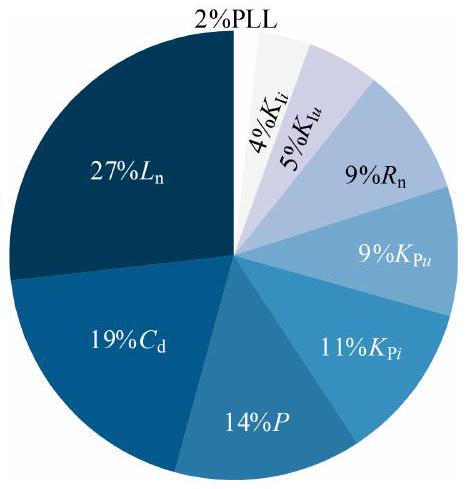}
\caption{Parameter-level sensitivity analysis of vehicle 4 for $\Lambda_{1}$ at 2.5 Hz in Case 1.}
\end{center}
\end{figure}

Furthermore, the infrastructure manager can share the admittance-level sensitivity function with the rolling stock supplier for further parameter-level sensitivity analysis. Based on the admittance-level sensitivity analysis results, the D-channel admittance of vehicle 1 can be reshaped to improve system stability by adding virtual impedance control [32]. Then the parameter-level sensitivity analysis can be implemented to identify the influence of circuit or control parameters on critical eigenlocus.

\section*{E. Parameter-Level Sensitivity Analysis}
Based on the port and admittance level sensitivity results, it can be found that vehicles 2 and 4 has the most influence on the system stability. Therefore, the system stability can be improved by optimizing the parameters of the vehicles 2 and 4. The parameter-level sensitivity analysis of eigenloci $\Lambda_{1}$ is performed by (20), which helps identify the critical parameters within the vehicle. The results of the parameter-level sensitivity analysis

\begin{figure}[H]
\begin{center}
  \includegraphics[alt={},max width=\columnwidth]{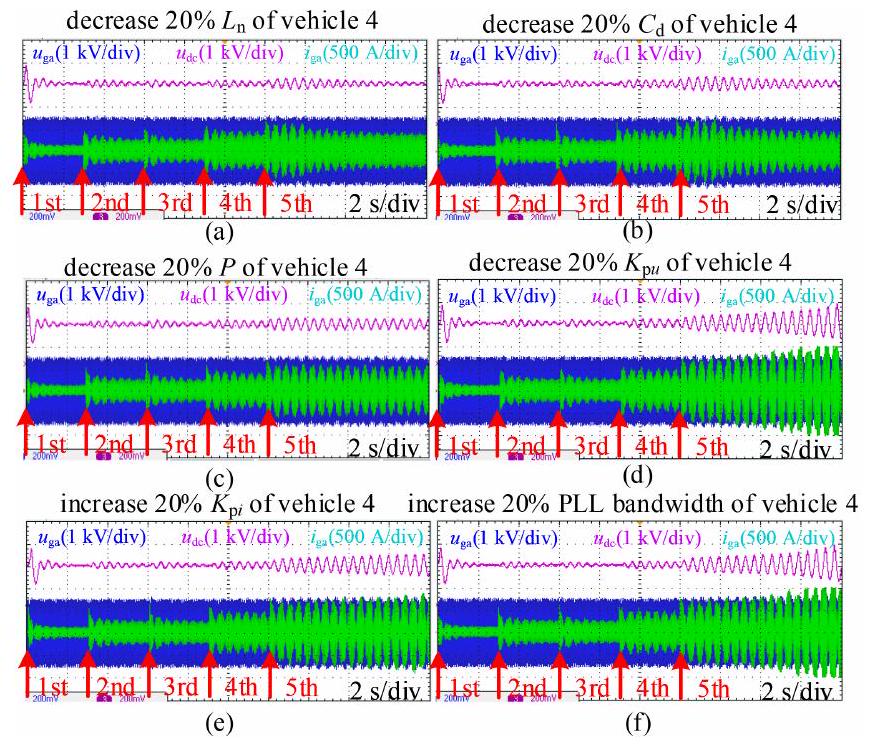}
\caption{Waveforms of line-side voltage and current, and dc-link voltage of vehicle. (a) Case 1.1. (b) Case 1.2. (c) Case 1.3. (d) Case 1.4. (e) Case 1.5. (f) Case 1.6.}
\end{center}
\end{figure}

are presented in a pie chart in Fig. 14, where the proportion of each parameter indicates its sensitivity to the system eigenlocus $\Lambda_{1}$ of vehicle 4 at 2 Hz in Case 1.

The analysis reveals that the on-board transformer parameters exhibit the highest sensitivity, accounting for $27 \%$ of the total. Following that, the dc-side capacitance demonstrates a sensitivity of $19 \%$. Among the control parameters, the proportional coefficient gain of the voltage outer loop and the current inner loop exhibit significant influence on the system, while the PLL bandwidth has the least impact. Overall, the eigenlocus $\Lambda_{1}$ is more sensitive to circuit parameters than control parameters. Therefore, to enhance the stability of the vehicle-grid system in the low-frequency range, priority should be given to designing and optimizing the circuit parameters of the vehicle.

To validate the accuracy of the parameter-level sensitivity results, additional cases (Cases 1.1-1.6) are designed based on Case 1, where the parameters of vehicle 4 are modified to observe their influence on system stability. The details of each case are as follows.

\begin{enumerate}
  \item Case 1.1: $20 \%$ decrease in the equivalent inductance ( $L_{\mathrm{n}}$ ) of the onboard transformer of vehicle 4 compared with Case 1.
  \item Case 1.2: 20\% decrease in the dc-side capacitance ( $C_{\mathrm{d}}$ ) of vehicle 4 compared with Case 1.
  \item Case 1.3: 20\% decrease in the operation power ( $P$ ) of vehicle 4 compared with Case 1.
  \item Case 1.4: 20\% decrease in the proportional coefficient gain of voltage control loop ( $K_{\mathrm{P} u}$ ) of vehicle 4 compared with Case 1.
  \item Case 1.5: 20\% increase in the proportional coefficient gain of current control loop ( $K_{\mathrm{P} i}$ ) of vehicle 4 compared with Case 1.
  \item Case 1.6: $20 \%$ increase in the PLL bandwidth of vehicle 4 compared with Case 1.
\end{enumerate}

\begin{figure}[H]
\begin{center}
  \includegraphics[alt={},max width=\columnwidth]{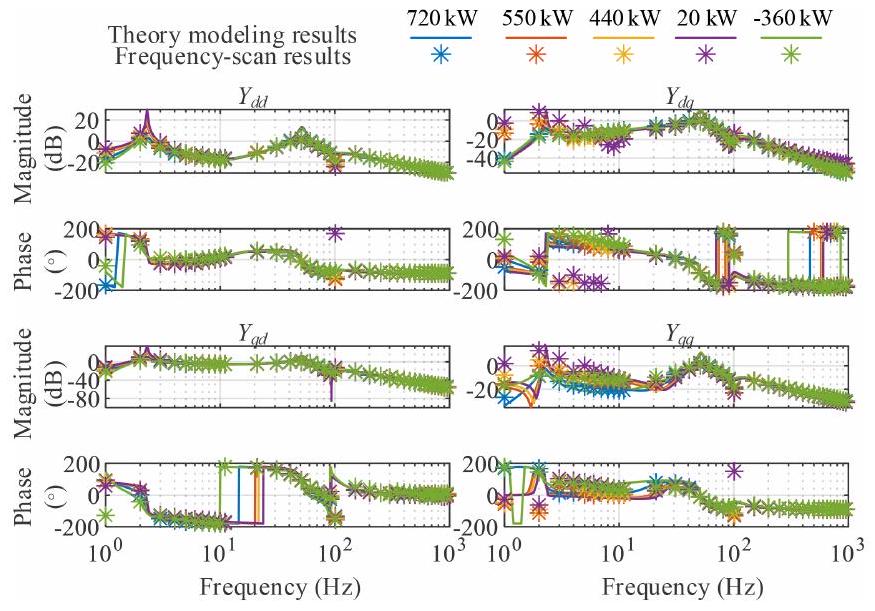}
\caption{DQ admittance for the vehicle at various operation power.}
\end{center}
\end{figure}

Fig. 16 presents the results of Cases 1.1-1.6. It can be observed that in Fig. 16(a) and (b), significant enhancements in system stability are observed by adjusting the parameters of the onboard transformer equivalent inductance ( $L_{\mathrm{n}}$ ) and the dc-side capacitance ( $C_{\mathrm{d}}$ ) of vehicle 4 . In addition, Fig. 16(b) shows that the system requires more time to reach a stable state after all vehicles are connected compared with Fig. 16(a), indicating a greater impact of the onboard transformer equivalent inductance $\left(L_{\mathrm{n}}\right)$ on system stability. After modifying the parameters $P, K_{\mathrm{P} u}$, and $K_{\mathrm{P} i}$, as shown in Fig. 16(c)-(e), the system maintains a critical stable state after all vehicles are connected, and voltage and current are oscillated at the PCC point. This suggests that these three parameters have a minor influence on system stability. In Fig. 16(f), altering the PLL bandwidth of vehicle 4, the system still exhibits amplification of oscillations. The waveforms obtained in Fig. 16(f) indicates that adjusting the PLL bandwidth has minimal effect on system stability. In conclusion, the results of Cases 1.1-1.6 align with the parameter-level sensitivity results, thus validating the accuracy of the approach.

\section*{V. Conclusion}
This article aims to analyze the stability of a multivehicleaccessed railway vehicle-grid system and identify the impact of each vehicle and its internal parameters on system stability. To achieve this, a multilevel frequency-domain sensitivity analysis method based on the CCM approach is proposed, along with a system-level compatibility test for railway vehicle-grid systems. The proposed sensitivity analysis method enables the evaluation of system stability from the port level to the parameter level, providing comprehensive insights into system stability analysis and controller design. Moreover, the proposed method is suitable for black-box systems and has lower computational complexity.

To validate the accuracy of the proposed method under different spatiotemporal distributions of vehicles, case studies of the vehicle-grid system are conducted based on actual train operation schedules. Experimental results demonstrate the precise assessment of system stability by the proposed vehicle-grid system method, as well as the accurate identification of the\\
dominant vehicle causing system oscillations by the sensitivity analysis method. Furthermore, it is observed that the electrical parameters of the vehicles have a more significant impact on the low-frequency stability of the system compared to the control parameters.

\section*{Appendix}
The DQ admittance of the vehicle is provided in Fig. 17, which shows the admittance results under the different operation power. The detailed modeling process and frequency-scan method of vehicle can be found in [20] and [33].

\begin{table}[H]
\centering
\caption{Nodes Voltage and Phase of Different Vehicles in Case 1}
\label{tab:nodes_voltage_case1}
\scriptsize
\setlength{\tabcolsep}{3pt}
\renewcommand{\arraystretch}{1.08}
\begin{tabular}{@{}lcc@{}}
\toprule
Vehicle number & Magnitude $V$ (p.u.) & Angle $\theta$ ($^\circ$) \\
\midrule
Vehicle 1 ($v_1$) & 0.98 & -0.9 \\
Vehicle 2 ($v_2$) & 0.96 & -3.2 \\
Vehicle 3 ($v_3$) & 0.95 & -4.4 \\
Vehicle 4 ($v_4$) & 0.97 & -1.8 \\
Vehicle 5 ($v_5$) & 0.98 & -3.6 \\
\bottomrule
\end{tabular}
\end{table}

\begin{table}[H]
\centering
\caption{Nodes Voltage and Phase of Different Vehicles in Case 2}
\label{tab:nodes_voltage_case2}
\scriptsize
\setlength{\tabcolsep}{3pt}
\renewcommand{\arraystretch}{1.08}
\begin{tabular}{@{}lcc@{}}
\toprule
Vehicle number & Magnitude $V$ (p.u.) & Angle $\theta$ ($^\circ$) \\
\midrule
Vehicle 1 ($v_1$) & 0.95 & -0.6 \\
Vehicle 2 ($v_2$) & 0.93 & -1.7 \\
Vehicle 4 ($v_4$) & 0.92 & -2.4 \\
Vehicle 5 ($v_5$) & 0.95 & -2.1 \\
Vehicle 6 ($v_6$) & 0.94 & -5.7 \\
\bottomrule
\end{tabular}
\end{table}

\section*{References}
[1] "National railway administration of the people's Republic of China." Accessed: Jan. 26, 2024. [Online]. Available: \href{http://www.nra.gov.cn/}{http://www.nra.gov.cn/}\\[0pt]
[2] H. Hu, H. Tao, F. Blaabjerg, X. Wang, Z. He, and S. Gao, "Train and stability evaluation in high-speed railways: Phenomena and modeling," IEEE Trans. Power Electron., vol. 33, no. 6, pp. 4627-4642, Jun. 2018.\\[0pt]
[3] Railway Applications-Fixed Installations and Rolling Stock-Technical Criteria for the Coordination Between Electric Traction Power Supply Systems and Rolling Stock to Achieve Interoperability, CENELEC Standard EN 50388-1, 2022.\\[0pt]
[4] H. Zhang, Z. Liu, S. Wu, and Z. Li, "Input impedance modeling and verification of single-phase voltage source converters based on harmonic linearization," IEEE Trans. Power Electron., vol. 34, no. 9, pp. 8544-8554, Sep. 2019.\\[0pt]
[5] Y. Zhou, H. Hu, X. Yang, Z. Meng, and Z. He, "Impacts of quadrature signal generation-based PLLS on low-frequency oscillation in an electric railway system," IEEE Trans. Transp. Electrific., vol. 7, no. 4, pp. 3124-3136, Dec. 2021.\\[0pt]
[6] Y. Liao, Z. Liu, G. Zhang, and C. Xiang, "Vehicle-grid system modeling and stability analysis with forbidden region-based criterion," IEEE Trans. Power Electron., vol. 32, no. 5, pp. 3499-3512, May 2017.\\[0pt]
[7] X. Lv, X. Wang, Y. Che, and R. Fu, "Eigenvalue-based harmonic instability analysis of electrical railway vehicle-network system," IEEE Trans. Transp. Electrific., vol. 5, no. 3, pp. 727-744, Sep. 2019.\\[0pt]
[8] H. Tao, H. Hu, X. Wang, F. Blaabjerg, and Z. He, "Impedance-based harmonic instability assessment in a multiple electric trains and traction network interaction system," IEEE Trans. Ind. Appl., vol. 54, no. 5, pp. 5083-5096, Sep./Oct. 2018.\\[0pt]
[9] G. Rogers, Power System Oscillations. Boston, MA, USA: Springer, 2000.\\[0pt]
[10] Y. Zhu, Y. Gu, Y. Li, and T. Green, "Participation analysis in impedance models: The grey-box approach for power system stability," IEEE Trans. Power Syst., vol. 37, no. 1, pp. 343-353, Jan. 2022.\\[0pt]
[11] Y. Liao, X. Wang, and X. Wang, "Frequency-domain participation analysis for electronic power systems," IEEE Trans. Power Electron., vol. 37, no. 3, pp. 2531-2537, Mar. 2022.\\[0pt]
[12] D. Yang and Y. Sun, "SISO impedance-based stability analysis for systemlevel small-signal stability assessment of large-scale power electronicsdominated power systems," IEEE Trans. Sustain. Energy, vol. 13, no. 1, pp. 537-550, Jan. 2022.\\[0pt]
[13] Y. Deng and Z. Lin, "Challenges and countermeasures facing the electrification project of Sichuan-Tibet railway," Electric Railway, vol. 30, no. 1, pp. 5-11, Mar. 2019.\\[0pt]
[14] Y. Deng, Z. Liu, Y. Wang, and K. Huang, "Study on application of allparallel DN power supply mode on montanic electrified railway," in Proc. Int. Conf. Electr. Inf. Technol. Rail Transp., 2016, pp. 153-162.\\[0pt]
[15] Z. Liu, G. Zhang, and Y. Liao, "Stability research of high-speed railway emus and traction network cascade system considering impedance matching," IEEE Trans. Ind. Appl., vol. 52, no. 5, pp. 4315-4326, Sep./Oct. 2016.\\[0pt]
[16] Y. Song, Z. Liu, A. Rønnquist, P. Nåvik, and Z. Liu, "Contact wire irregularity stochastics and effect on high-speed railway pantograph-catenary interactions," IEEE Trans. Instrum. Meas., vol. 69, no. 10, pp. 8196-8206, Oct. 2020.\\[0pt]
[17] J. A. Aguado, A. J. S. Racero, and S. de la Torre, "Optimal operation of electric railways with renewable energy and electric storage systems," IEEE Trans. Smart Grid, vol. 9, no. 2, pp. 993-1001, Mar. 2018.\\[0pt]
[18] L. Luo and S. V. Dhople, "Spatiotemporal model reduction of inverterbased islanded microgrids," IEEE Trans. Energy Convers., vol. 29, no. 4, pp. 823-832, Dec. 2014.\\[0pt]
[19] A. Rygg and M. Molinas, "Apparent impedance analysis: A small-signal method for stability analysis of power electronic-based systems," IEEE Trans. Emerg. Sel. Topics Power Electron., vol. 5, no. 4, pp. 1474-1486, Dec. 2017.\\[0pt]
[20] Y. Liao, Z. Liu, X. Hu, and B. Wen, "A DQ-frame impedance measurement method based on Hilbert transform for single-phase vehicle-grid system," in Proc. IEEE Transp. Electrif. Conf. Expo, 2017, pp. 1-6.\\[0pt]
[21] K. Mongkoldee and T. Kulworawanichpong, "Current-based NewtonRaphson power flow calculation for AT-fed railway power supply systems," Int. J. Elect. Power Energy Syst, vol. 98, pp. 11-22, Jun. 2018.\\[0pt]
[22] W. Cao, Y. Ma, L. Yang, F. Wang, and L. M. Tolbert, "D-Q impedance based stability analysis and parameter design of three-phase inverter-based ac power systems," IEEE Trans. Ind. Electron., vol. 64, no. 7, pp. 6017-6028, Jul. 2017.\\[0pt]
[23] W. Zhou, Y. Wang, R. E. Torres-Olguin, and Z. Chen, "DQ impedance reshaping of three-phase power-controlled grid-connected inverter for low-frequency stability improvement under weak grid condition," in Proc. IEEE Energy Convers. Congr. Expo., 2020, pp. 1678-1685.\\[0pt]
[24] J. Chen et al., "Analysis and control of cascaded energy storage system for energy efficiency and power quality improvement in electrified railways," IEEE Trans. Transp. Electrific., early access, Jun. 21, 2023, doi: 10.1109/TTE.2023.3287891.\\[0pt]
[25] N. Cifuentes, M. Sun, R. Gupta, and B. C. Pal, "Black-box impedancebased stability assessment of dynamic interactions between converters and grid," IEEE Trans. Power Syst., vol. 37, no. 4, pp. 2976-2987, Jul. 2022.\\[0pt]
[26] W. Zhou, R. E. Torres-Olguin, Y. Wang, and Z. Chen, "A gray-box hierarchical oscillatory instability source identification method of multiple-inverter-fed power systems," IEEE Trans. Emerg. Sel. Topics Power Electron., vol. 9, no. 3, pp. 3095-3113, Jun. 2021.\\[0pt]
[27] C. Zhang, M. Molinas, A. Rygg, J. Lyu, and X. Cai, "Harmonic transfer-function-based impedance modeling of a three-phase VSC for asymmetric AC grid stability analysis," IEEE Trans. Power Electron., vol. 34, no. 12, pp. 12552-12566, Dec. 2019.\\[0pt]
[28] Z. Li, M. Zhu, C. Hou, H. Wang, Y. Li, and X. Cai, "Impedance modelling mechanisms and stability issues of single phase inverter with SISO structure and frequency coupling effect," IEEE Trans. Energy Convers., vol. 37, no. 1, pp. 573-584, Mar. 2022.\\[0pt]
[29] X. Meng et al., "Conversion and SISO equivalence of impedance model of single-phase converter in electric multiple units," IEEE Trans. Transp. Electrific., vol. 9, no. 1, pp. 1363-1378, Mar. 2023.\\[0pt]
[30] H. Wang, W. Mingli, and J. Sun, "Analysis of low-frequency oscillation in electric railways based on small-signal modeling of vehicle-grid system in DQ frame," IEEE Trans. Power Electron., vol. 30, no. 9, pp. 5318-5330, Sep. 2015.\\[0pt]
[31] H. Lin, S. Niu, Z. Xue, and S. Wang, "A simplified virtual-vectorbased model predictive control technique with a control factor for threephase SPMSM drives," IEEE Trans. Power Electron., vol. 38, no. 6, pp. 7546-7557, Jun. 2023.\\[0pt]
[32] J. Fang, X. Li, H. Li, and Y. Tang, "Stability improvement for three-phase grid-connected converters through impedance reshaping in quadratureaxis," IEEE Trans. Power Electron., vol. 33, no. 10, pp. 8365-8375, Oct. 2018.\\[0pt]
[33] Y. Liao, Z. Liu, H. Zhang, and B. Wen, "Low-frequency stability analysis of single-phase system with DQ-frame impedance approach: Impedance modeling and verification," IEEE Trans. Ind. Appl., vol. 54, no. 5, pp. 4999-5011, Sep./Oct. 2018.\\

\end{document}